\documentclass[12pt]{aastex}
\usepackage{natbib}
\newcommand{\dfrac}[2]{{\displaystyle \frac{#1}{#2}}  }

\newcommand{\eqref}[1]{(\ref{#1})}

\shortauthors{Grady et al.}
\shorttitle{Spiral Arms in MWC 758}

\begin{document}
\title{Spiral Arms in the Asymmetrically Illuminated Disk of MWC 758 and Constraints on Giant Planets}
\author {C.A. Grady\altaffilmark {1,2}, T. Muto\altaffilmark {3},  J. Hashimoto\altaffilmark {4}, M. Fukagawa\altaffilmark{5}, T. Currie\altaffilmark{6, 7},  B. Biller\altaffilmark {8},
C. Thalmann\altaffilmark{9}, M. L. Sitko\altaffilmark{10},  R. Russell\altaffilmark{11}, J. Wisniewski\altaffilmark{12}, R. Dong\altaffilmark{13}, J.  Kwon\altaffilmark{4,14}, S. Sai\altaffilmark{5}, J. Hornbeck\altaffilmark{15}, G. Schneider\altaffilmark{16}, D. Hines\altaffilmark{17}, 
 A. Moro Mart\'in\altaffilmark{18}, M. Feldt\altaffilmark {8}, Th. Henning\altaffilmark{8},  J.-U. Pott\altaffilmark{8}, M. Bonnefoy\altaffilmark{8},  J. Bouwman\altaffilmark{8}, S. Lacour\altaffilmark{19}, A. Mueller\altaffilmark{8}, A. Juh\'asz\altaffilmark{20}, 
A. Crida\altaffilmark{21}, G. Chauvin\altaffilmark{22}, S. Andrews\altaffilmark{23}, D. Wilner\altaffilmark{23}, A. Kraus\altaffilmark{24}, 
S. Dahm\altaffilmark{25}, T. Robitaille\altaffilmark{23}, H. Jang-Condell\altaffilmark{26}, L. Abe\altaffilmark {27},  E. Akiyama\altaffilmark{4},  W. Brandner\altaffilmark{8}, T. Brandt\altaffilmark{13},
J. Carson\altaffilmark{28}, S. Egner\altaffilmark{29}, K. B. Follette\altaffilmark{16},  M. Goto\altaffilmark{30}, O. Guyon\altaffilmark{31}, Y. Hayano\altaffilmark{29}, M. Hayashi\altaffilmark{4},
S. Hayashi\altaffilmark{29}, K. Hodapp\altaffilmark{31}, M. Ishii\altaffilmark{29}, M. Iye\altaffilmark{4}, M. Janson\altaffilmark{13}, R. Kandori\altaffilmark{4},
G. Knapp\altaffilmark{13}, T. Kudo\altaffilmark{29}, N. Kusakabe\altaffilmark{4}, M. Kuzuhara\altaffilmark{4,32}, S. Mayama\altaffilmark{34}  M. McElwain\altaffilmark{6}, T. Matsuo\altaffilmark{4},
S. Miyama\altaffilmark{35}, J.-I. Morino\altaffilmark{4}, T. Nishimura\altaffilmark{29}, T.-S. Pyo\altaffilmark{29}, G. Serabyn\altaffilmark{36}, H. Suto\altaffilmark{4}, 
R. Suzuki\altaffilmark{4}, M. Takami\altaffilmark{37}, N. Takato\altaffilmark{30}, H. Terada\altaffilmark{29},  D. Tomono\altaffilmark{30}, E. Turner\altaffilmark{12,31},
M. Watanabe\altaffilmark{38}, T. Yamada\altaffilmark{39}, H. Takami\altaffilmark{4}, T. Usuda\altaffilmark{29}, and 
M. Tamura\altaffilmark{4,14}} 

\altaffiltext {1} {Eureka Scientific, 2452 Delmer, Suite 100, Oakland CA 96002, and ExoPlanets and Stellar Astrophysics Laboratory, Code 667, Goddard Space Flight Center, Greenbelt, MD 20771 USA, carol.a.grady@nasa.gov} 
\altaffiltext {2} {Goddard Center for Astrobiology}
\altaffiltext {3} {Division of Liberal Arts, Kogakuin University, 1-24-2, Nishi-Shinjuku, Shinjuku-ku, Tokyo, 163-8677, Japan}
\altaffiltext {4} {National Astronomical Observatory of Japan, 2-21-1 Osawa, Mitaka, Tokyo 181-8588, Japan}
\altaffiltext {5} {Department of Earth and Space Science, Graduate School of Science, Osaka University, 1-1, Machikaneyama, Toyonaka, Osaka 560-0043, Japan}
\altaffiltext {6} {ExoPlanets and Stellar Astrophysics Laboratory, Code 667, NASA's  Goddard Space Flight Center, Greenbelt, MD 20771 USA}
\altaffiltext{7}{University of Toronto, Department of Astronomy and Astrophysics}
\altaffiltext {8} {Max Planck Institute for Astronomy, Heidelberg, Germany}
\altaffiltext {9} {Astronomical Institute \char'134 Anton Pannekoek", University of Amsterdam, Science Park 904, 1098 XH Amsterdam, The Netherlands }
\altaffiltext {10} {Space Science Institute, 4750 Walnut St., Suite 205, Boulder, CO 80301, USA ; Department of Physics, University of Cincinnati, Cincinnati, OH 45221-0011, USA ; Visiting Astronomer, NASA Infrared Telescope Facility, operated by the University of Hawaii under contract to NASA. }
\altaffiltext {11} {The Aerospace Corporation, Los Angeles, CA 90009, USA ; Visiting Astronomer, NASA Infrared Telescope Facility, operated by the University of Hawaii under contract to NASA. }
\altaffiltext {12} {Homer L. Dodge Department of Physics and Astronomy, The University of Oklahoma, Norman, OK 73019, USA}
\altaffiltext {13} {Department of Astrophysical Sciences, Princeton University, NJ08544, USA}
\altaffiltext {14} {Department of Astronomical Science, The Graduate University for Advanced Studies~(SOKENDAI), 2-21-1 Osawa, Mitaka, Tokyo 181-8588, Japan}
\altaffiltext{15}{Department of Physics and Astronomy, University of Louisville, Louisville, KY 40292, USA}
\altaffiltext{16}{Steward Observatory, The University of Arizona, Tucson, AZ  85721-0065, USA}
\altaffiltext{17} {Space Telescope Science Institute, 3700 San Martin Drive, Baltimore, MD 21218, USA} 
\altaffiltext{18} {Departamento de Astrofisica, CAB (INTA-CSIC), Instituto Nacional de TŽcnica Aeroespacial, Torrej—n de Ardoz, 28850 Madrid, Spain } 
\altaffiltext{19} {LESIA-Observatoire de Paris, CNRS, UPMC Univ. Paris 06, Univ. Paris-Diderot, 92195, Meudon, France}
\altaffiltext{20} {Leiden Observatory, Leiden University, PO Box 9513, NL-2300 RA Leiden, The Netherlands}
\altaffiltext{21} {Laboratoire Lagrange, UMR7293, UniversitŽ de Nice Sophia-Antipolis, CNRS, Observatoire de la C™te d'Azur, Boulevard de l'Observatoire, 06304 Nice Cedex 4, France}
\altaffiltext{22}{Laboratoire d'Astrophysique de l'Observatoire de Grenoble, 414, Rue de la Piscine, Domaine Universitaire, BP 53, 38041 Grenoble Cedex 09, France}
\altaffiltext{23}{Harvard-Smithsonian Center for Astrophysics, 60 Garden Street, Cambridge, MA 02138, USA}
\altaffiltext{24} {Institute for Astronomy, University of Hawaii, 2680 Woodlawn Drive, Honolulu, HI 96822, USA ; Hubble Fellow}
\altaffiltext{25} {W. M. Keck Observatory, 65-1120 Mamalahoa Hwy, Kamuela, HI 96743, USA}
\altaffiltext{26} {Department of Physics \& Astronomy, University of Wyoming, Laramie, WY 82071, USA}
\altaffiltext{27}{Laboratoire Lagrange, UMR7293, UniversitŽ Le de Nice-Sophia Antipolis, CNRS, Observatoire de la C™te d'fAzur, 06300 Nice, France }
\altaffiltext{28} {Department of Physics \& Astronomy,  The College of Charleston. Charleston, SC 29424, USA} 
\altaffiltext{29}{Subaru Telescope, 650 North AÕohoku Place, Hilo, HI 96720, USA,}
\altaffiltext{30}{Universit\"ats-Sternwarte M\"unchen Scheinerstr. 1, D-81679 Munich, Germany}
\altaffiltext{31}{The Kavli Institute for the Physics and Mathematics of the Universe, The University of Tokyo, Kashiwa 227-8568, Japan}
\altaffiltext{32}{ Institute for Astronomy, 640 N. A'ohoku Place, Hilo, HI 96720,  USA}
\altaffiltext{33}{Department of Earth and Planetary Science, University of Tokyo, 7-3-1 Hongo, Tokyo 113-0033, Japan}
\altaffiltext{34}{The Graduate University for Advanced Studies(SOKENDAI), Shonan International Village, Hayama-cho, Miura-gun, Kanagawa 240-0193, Japan} 
\altaffiltext{35}{Office of the President, Hiroshima University, 1-3-2 Kagamiyama, Higashi-Hiroshima, JAPAN,739-8511}
\altaffiltext{36}{Jet Propulsion Laboratory, M/S 171-113 , 4800 Oak Grove Drive,  Pasadena, CA 91109, USA}
\altaffiltext{37}{Institute of Astronomy and Astrophysics, Academia Sinica, P.O. Box 23-141, Taipei 10617, Taiwan, R.O.C}
\altaffiltext{38}{Department of Cosmosciences, Hokkaido University, Sapporo 060-0810, Japan } 
\altaffiltext{39}{Astronomical Institute, Tohoku University, Aoba, Sendai 980-8578, Japan}
\begin{abstract}
We present the first near-IR  scattered light detection of the transitional disk associated with the Herbig Ae star MWC 758   using data obtained
as part of the {\it Strategic Exploration of Exoplanets and Disks with Subaru}, and  1.1$\mu$m   HST/NICMOS  data.  While sub-millimeter studies suggested there is a  dust-depleted cavity with r=0\farcs35,  we find scattered light  as close as 0.1\arcsec  ~(20-28 AU) from the star, with no visible cavity at H, K', or K$_s$. We  find two small-scaled spiral structures  which asymmetrically shadow the outer disk.   We model one of  the spirals using  spiral density wave theory, and derive a disk aspect ratio of h$\sim$0.18, indicating a dynamically
warm disk. If  the spiral pattern is excited by a perturber,  we estimate its mass to be 5$^{+3}_{-4}$ M$_{J}$, in the range where planet filtration models predict 
accretion continuing onto the star.  Using a combination of non-redundant aperture masking data at L$'$ and angular differential imaging
with Locally Optimized Combination of Images at K$'$ and K$_s$,  we exclude stellar or massive brown dwarf companions within 300 mas of the Herbig Ae star,
and all but planetary mass companions exterior to 0\farcs5.  We reach 5-$\sigma$ contrasts limiting companions to planetary masses, 3-4 M$_{J}$ at 1\farcs0 and
2 M$_{J}$ at 1\farcs55  using the COND models.  Collectively, these data strengthen the case for MWC 758
already being  a young planetary system. 
\end{abstract}
\keywords{circumstellar matter Ñ instrumentation: high angular resolution $Ñ$ polarization Ñ planetary systems: protoplanetary disks Ñ stars: individual   
 (MWC~758) $Ñ$ waves}

\section {Introduction}

We live in a planetary system with four  giant planets, and over the past 15 years a wealth of other massive planets have been identified from radial velocity, transit, and microlensing studies, as well as from direct imaging of nearby stars.  These studies have told us a great deal about the 
frequency of giant planets  around older, Main Sequence stars, but   do not provide insight into when, where, and how frequently they form in their natal circumstellar  disks, beyond the  trivial constraint that they must have formed while the disks retained abundant gas. A census of signatures of giant planet presence in young disks  can be used to bound both the time required to form such bodies, and the portions of the disk occupied by giant planets. Such data are needed to constrain mechanisms for their formation and migration. Interferometric detection of young, Jovian-mass exoplanets and brown dwarfs  within their disks is challenging, but an emerging field \citep []{Kraus12, Huelamo11}, however only a small portion of the disk is sampled. Search techniques focussing on the macroscopic, albeit indirect, signatures that such planets induce in their host disks offer an alternate way of carrying out such a giant planet census. 

Recent, high-contrast polarimetric differential imagery of the Herbig F star SAO 206462 has demonstrated  one such indirect signature: the presence of spiral arms interpreted as being excited by spiral density waves \citep []{Muto12}. For SAO 206462, 2 arms were clearly resolved from the disk, and have amplitudes consistent with perturbers that are $\sim$Saturn-mass objects. SAO 206462 is estimated to have an age of 9$\pm$2 Myr \citep []{Mueller11}, older than age estimates for the formation of Saturn \citep []{C-R09} and potentially Jupiter  in our Solar System. While tantalizing, these data, by themselves, do not provide insight into how common spiral features are in the disks of young stars, or at what point in the evolution of a disk they become detectable.  

The Herbig Ae star MWC 758 \citep [HD 36112, A8Ve; V=8.27, B-V=0.3; H=6.56][] {Besk99},  \citep[A5IVe][]{Meeus12} has also been identified as hosting a partially cleared cavity detected
in the sub-millimeter within 0\farcs 2   of the star \citep[]{Isella10, Andrews11}.  The star has been dated to 3.5$\pm$2 Myr \citep[]{Meeus12}, comparable to the era of giant planet formation in our Solar System. The disk still contains molecular and atomic gas \citep[]{Isella10, Salyk11, Meeus12}, while accretion continues onto the star \citep[]{Besk99}.  The revised Hipparcos parallax data place the star at 279
$^{+94}_{-58}$
pc \citep[]{vanLeeuwen07}, but much of the literature uses the older Hipparcos measurement of 200$^{+60}_{-40}$ pc \citep[]{vandena98}. To ensure compatibility with the older literature, we use
both distance estimates in this paper. 
 Like SAO 206462, the disk has a low inclination from pole-on \citep[i=21$\pm$2$^\circ$][]{Isella10},   facilitating  detection of structure in the inner parts of the system.  MWC 758 has a relatively small depletion of mm-sized dust grains in the
inner portions of its disk \citep[] {Andrews11}, and also has not been previously reported to host the distinctive mid-IR dip in the IR spectral energy distribution (SED)  \citep[]{Isella10}. Millimeter  interferometry traces gas up to 2\farcs6 from the star \citep[] {Chapillon08}, while millimeter dust emission can be traced $\sim$1\arcsec ~from the star.  The scale of the disk, the 
gap size, and the inclination of the outer disk all make this a suitable system to test the hypothesis that SAO 206462 is not unique in hosting spiral arms potentially associated with Jovian-mass planets. 

We report the first successful imaging of the disk of MWC 758 in scattered light, the discovery of spiral arms, the detection of a mid-IR dip in the IR SED
as well as apparent mid-IR photometric variability, and constrain the mass of companions in and immediately exterior to the disk 
 using a mixture of NIR high-contrast imagery, sparse-aperture masking interferometry at L', and the measured properties of the spiral arms. 

\section {Observations and Data Reduction}
In this section we discuss observations and data reduction for high-contrast imaging carried out using two instruments at the Subaru Telescope,  a re-analysis of HST
coronagraphic imagery, aperture polarimetry and photometry, and assembly of  IR spectral energy distribution data, and VLT sparse aperture masking interferometry. 
The high-contrast Subaru imagery is processed both for optimal disk detection and for giant planet searches. 

\subsection {Subaru/HiCIAO Polarimetric Differential Imagery}

MWC 758  was observed in the H-band ($\lambda$=1.635$^{+0.155}_{-0.145} ~ \mu$m) using the high-contrast imaging instrument HiCIAO \citep[]{Tamura06,  Hodapp08,  Suzuki10} on the Subaru Telescope on 2011 Dec. 24 UT as part of the  {\it Strategic Exploration of Exoplanets and Disks with Subaru} (SEEDS)
 program \citep[]{Tamura09}.  The AO 188 adaptive optics system  \citep []{Minowa10} provided a stable stellar point spread function (PSF, FWHM = 0\farcs07).  We observed with  a combined angular differential imaging (ADI)  \citep []{Marois2006} and polarization differential imaging (PDI) mode with a field of view of 10\arcsec  ~by 20\arcsec  ~and pixel scale of 9.5 mas pixel$^{-1}$.  A 0\farcs 3
-diameter circular occulting mask was used to suppress the bright stellar halo. Half-wave plates were placed at 4 angular positions from 0$^\circ$, 45$^\circ$, 22.5$^\circ$ and 67.5$^\circ$ in sequence with one 30 sec exposure per wave plate position for a total of 15 datasets, and 
a total integration time for the polarized intensity (PI) image of 1,440 s, after removing 3 low-quality images with FWHM$>$ 0\farcs1,  by careful inspections of the stellar PSF.  The field rotation in this dataset was 9$^\circ$ on the sky, too small for ADI reduction of the data.  While a PSF
star was observed immediately following the PDI+ADI image set, the PSF had changed significantly from that observed
during the observation of MWC 758, precluding conventional PSF subtraction. H-band photometry for MWC 758, taken with the telescope immediately before the PDI+ADI dataset yielded H = 6.$^m$48$\pm$0.$^m$01 in the MKO filter system. The polarimetric data were reduced using  the procedure  described in  \citet{hash11} using IRAF\footnote{
  IRAF is distributed by the National Optical Astronomy Observatories, which are operated
  by the Association of Universities for Research in Astronomy,
  Inc., under cooperative agreement with the National Science Foundation.}.   
    
 % For symmetric disks, this residual stellar halo is usually polarized along the disk
 % semi-minor axis (here PA=155$^\circ$/335$^\circ$), since the polarized fraction of the forward and
%  backscattering is smaller given a  deviation from 90$^{\circ}$ in the scattering angle.  Aligned polarization vectors have previously been observed  in polarimagery of GM Aur and LkCa 15 \citep{potter05}.
 
 % The subtraction process of the $``$polarized halo$"$ is as follows: 
%  First, we construct a number of polarized haloes 
%  by multiplying the added image of $o$- and $e$-rays by a variety of 
%  degrees of polarization (P) and polarimetric angles ($\theta$).
%  Using the results of aperture polarimetry for HiCIAO data is 
%  the straightforward method, but we did not observe unsaturated MWC 758 
%  without occulting mask with PDI. Therefore we tried to find the best polarized halo. 
%  Then we subtracted the polarized halo from obtained final Stokes $Q$ and $U$ images.
%  To determine which parameters of P and $\theta$ are better, 
%  we made histograms of angles between polarization vectors and lines 
%  from the mask center to the vector position. We finally concluded that the polarized halo
%  with P=0.14\% and $\theta=74^{\circ}$ offered the best minimization of 
%  non-centrosymmetric polarization vectors. The halo is not preferentially polarized along 
%  the disk semi-minor axis, which may indicate the presence of asymmetric structure in
%  the disk of MWC 758.   Figure 1 shows the results of the SEEDS polarization pipeline before and after correction for the
%  polarized halo. 

\subsection { Subaru/HiCIAO K$_s$ ADI observations} 

A second HiCIAO dataset was obtained as part of the SEEDS program at K$_s$ ($\lambda$=2.150$\pm$0.160 $\mu$m)  on 2011 Dec. 26 UT, with 17 minutes of integration time
in ADI mode. The field of view for this dataset was 20\arcsec ~x 20\arcsec .  The data were obtained in direct imaging mode, with no coronagraph, but with short 1.5s  frame  times to minimize image saturation. Only the peak of the PSF of MWC 758 was saturated in the K$_s$ data. 
The 1.5s integrations were 
coadded into frames of typically 20 integrations, but after culling, some 10-integration frames were also included. 
After culling of bad frames, 640s  of integration time were available.  The data
were processed using variants of  \textit{Locally Optimized Combination of Images} (LOCI) in two ways:  using aggressive LOCI (na = 300, nfwhm = 0.5) for a companion search and
conservative LOCI  (na =10,000, nfwhm = 1.5) to search for structure in the disk.  A discussion of these approaches
is in Thalmann et al. (2010).  Either LOCI variant suppresses azimuthally-symmetric or large angle structure, resulting
in voids compared to either the PI data or PSF-subtracted imagery.  

\subsection{Subaru/IRCS K$'$ Data}
MWC 758 and a PSF reference star (HD 242067) were imaged on November 10, 2011 with the Subaru Telescope 
using the Infrared Camera and Spectrograph \citep[IRCS;][]{Tokunaga1998} 
and  AO188 in natural guide star mode as part of program S11B-012. 
The data were obtained in the Mauna Kea K$'$ filter 
($\lambda$ = 2.12 $\mu m$).  
The native pixel scale is 20.57 mas/pixel; we used the 0.15\arcsec{} 
diameter coronagraphic mask to reduce light from the PSF core spilling 
out to angular separations of interest for  disk detection.
Despite variable weather conditions, 
AO-188 delivered corrected images with a full-width half-maximum of 
0\farcs07.  

Our data consist of coadded 15-second exposures taken in ADI  mode \citep{Marois2006} through transit HA = [-0.6,0.9] with 
a total field rotation of 127$^{\circ}$.  Our cumulative integration time for MWC 758 
and HD 242067 is 38 minutes and 10 minutes, respectively.  
Basic image processing follows steps outlined in \citet{Currie2011a,Currie2011b} for IRCS, including 
dark subtraction, flatfielding, and distortion correction (corrected pixel scale = 20.53 mas/pixel).  
We considered all HD 202067 frames to construct a reference PSF.

\textbf{Disk Imaging:} To extract a detection of the MWC 758 disk, we explored two approaches: one using PSF subtraction of MWC 758 
frames by a median HD 242067 frame and one 
leveraging upon an ``adaptive" LOCI approach  \citep[A-LOCI, ][]{Currie2012} 
for this subtraction.  In both cases, we first identified the highest-quality images of MWC 758.  
Specifically, we monitored 
the brightness of a background star 2\arcsec{} from MWC 758 ($\sim$ 15 magnitude) and selected only 
the frames where the star's brightness is within 20\% of the peak brightness of any frame.  
This selection criterion identified 50 MWC 758 frames (750 s total) to focus on for PSF subtraction.
Further reduction steps are as follows:

\begin{itemize}
\item \textit{Classical PSF Subtraction} -- Here, we subtract the median-combined HD 242067 image from 
each MWC 758 image.  We choose a weighting that minimizes the residuals between separations of 50 pixels 
and 150 pixels ($\sim$ 1\arcsec{}--3\arcsec{}), though weightings over slightly different ranges in 
angular separations (i.e. 0\farcs75--2\farcs5) achieve similar results.  We also performed 
a second subtraction weighting the reference PSF by an additional 10\% above the value that 
minimized the residuals for each MWC 758 image.  We then derotated and median-combined each PSF-subtracted 
MWC 758 image.
\item \textit{A-LOCI} -- Here, we treat all HD 242067 images separately as a ``reference PSF library".  
Instead of subtracting each reference image from the MWC 758 image at once, we follow the (A)-LOCI approach, 
subdividing each science image and reference image into annular sections \citep[see ][]{Currie2012,  Lafreniere2007} and 
determining coefficients for each reference PSF section that minimize the subtraction residuals within that 
section for a given science image.  To better ensure that any ``disk" signal is not instead due to PSF mismatch, 
we compute the cross-correlation function, $r_\mathrm{corr}$, for each pair of science-reference images and remove poorly correlated 
image pairs (we find the best results for $r_\mathrm{corr}$ $>$ 0.85).  We set the \textit{optimization} area (in units of the 
image FWHM) over which we define the coefficients to $N_\mathrm{A}$ = 3000 and adopt a radial width for the 
\textit{subtraction} annulus of $dr$ = 140, equal roughly to the angular extent of the disk as seen in the submm 
\citep[e.g.][]{Andrews11}\footnote{We detect the inner disk spirals as long as we set $dr$ to be larger 
than the submm-resolved disk gap ($dr$ $>$ 25)}. 
\end{itemize}  

\textbf{A-LOCI Limits on Planetary Companions} -- To search for companions to MWC 758, we employed 
much more aggressive PSF subtraction settings using A-LOCI on the full set of images 
(150 images; 37.5 minutes).  We adopted A-LOCI settings 
of $\delta$ = 0.70, $N_{A}$ = 250, $dr$ = 5, and $r_\mathrm{corr}$ = 0.95 \citep[see ][]{Currie2012, Lafreniere2007}.  
Additionally, we employed a moving-pixel mask over the subtraction zone, using only the 
pixels outside of this zone to determine the LOCI coefficients to reconstruct a reference PSF 
annular region.
We then quantify and correct for self-subtraction inherent in A-LOCI by 
comparing the flux of fake point sources prior to and after processing. 
We use the measured brightness of the  star visible at 2\arcsec{} ($K^\prime$ $\sim$ 16.93) 
and assume a brightness for MWC 758 of m($K^\prime$) = 5.804 
for flux calibration and thus to determine the planet-to-star contrast. 

\subsection{Archival HST Coronagraphic Imagery}

MWC 758 was observed in the broadband optical  twice by HST/STIS  on 2000 Jan. 16 and Jan. 19 as part of HST-GO-8474,
with the star placed
at a point along the STIS corongraphic wedge structure (wedgeA1.0) where the wedge occulted the inner 0\farcs5. 
 The STIS data (observation ids=O5KQ3010-20 and O5KQ0410-20) were a poor color match to the available library of PSF
template data \citep []{Grady05},  yielding disk non-detections with large color mis-match errors. When roll-differenced \citep[]{Lowrance05},
a null detection was also seen, indicating either nebulosity which was azimuthally-symmetric over 15$^\circ$, or concentrated within
0\farcs5 of the star.  The STIS data indicate only 3 point sources in the 25\arcsec x 50\arcsec ~field other
than MWC 758,  with the nearest at a separation of 2\farcs1 at
PA=309$^\circ$. 

MWC 758 was also observed by HST/NICMOS as part of HST-GO-10177 on 2005 Jan. 7.  The observations consisted of  two independent target acquisitions, 
short direct images in F171M,  the coronagraphic observations at F110W ($\lambda_{cent}$=1.1$\pm$0.3 $\mu$m) with the region
at r$\leq$0\farcs3 occulted, and short direct-light F110W imagery with the star unocculted
before rolling HST by 29.9$^\circ$ and repeating the sequence.  The direct-light imagery yielded F110W magnitudes of 7.$^m$314$\pm$0.014. 
The observations obtained  at a given spacecraft orientation constitute  a \char'134 visit". 
The coronagraphic data for each visit were reduced as described in  \citet []{Schneider06}, and then  a suite of PSF template data
from the same HST cycle were scaled, registered, and subtracted from the MWC 758 imagery. 
For the first visit for MWC 758 (visit 31), PSF template data from visits 3D (HD 15745, visit 2), visit 62 and 61(HD 83870), visit 53
(HD 35841 visit 1), visit 41 (HD 22128, visit 1), and 6B (HD 204366, visit 1) provided good color and HST residual
matches to MWC 758. For the second visit (visit 32), data from visit 48 (HD 38207 visit 1), 4C (HD 30447, visit 1),4D and 4E (HD 72390, visit 1and 2),
42 (HD 22128 visit 2), 62 (HD 83870 visit 2), 65 and 66 (HD 164249, visits 1 and 2) , and 6B (HD 204366, visit 1)  provided the best matches. The
use of independent target acquisitions and largely  independent suites of PSF template data mean that medians of the net images
for each visit provide two essentially independent observation sets for MWC 758.  Both final images  provide no indication of nebulosity
for r$\geq$0\farcs8, with  the bulk of the detected signal within 0\farcs5 of MWC 758.   

The angular scale of the nebulosity is
too small compared to the NICMOS F110W resolution element to assess azimuthal symmetry. but we have carried out aperture
photometry between 0\farcs3$<$r$<$0\farcs5  for each of the final net visit images. The total
brightness in those annuli in instrumental counts/s/pixel are 5183 for visit 31 and 5419 for visit 32, yielding only a $\pm$2\% peak to peak
difference about a mean of 5301 counts/s/pixel. Using the NICMOS  photometric calibration of 1c/s/pixel=1.26$\times$10$^{-6}$ Jy, the total F110W flux
density in the measurement annulus is 6.7 mJy.  The 2MASS J magnitude for MWC 758 is 7.22 (2.07Jy), so the total 1.1$\mu$m 
light scattered by this portion of the disk is 0.32\%.  The nearby point source  noted in the 2000 STIS imagery is recovered at r=2\farcs215, PA=310.86$^\circ$. It is not co-moving with MWC 758, and thus is a background object. 

\subsection {Infrared Survey Facility/Simultaneous IR polarimeter (SIRPOL) Aperture Polarimetry and Photometry} 
NIR aperture photometry and polarimetry was carried out for MWC 758 on 8 March 2012
using SIRPOL \citep []{Kandori06} at the IRSF.  The SIRPOL observations were not accompanied by observation of a polarimetric standard star, so the 2MASS system was used to calibrate the NIR magnitudes of MWC 758 using field stars. 
We measured J=7.$^m$1.93, H=6.$^m$560, Ks=5.$^m$804 with polarizations of 1.63\% at J, 1.60\% at H and 0.81\% at Ks.  The polarization levels are typical of low-inclination young stellar objects \citep []{Pereyra09}. 

\subsection {The IR Spectral Energy Distribution of MWC 758}

MWC 758 is a well-studied Herbig Ae star with data from  the far-UV through the millimeter. We have assembled a composite SED including 
IUE spectra SWP 53939 and LWP 30023, optical through M-band photometry  \citep[] {Malfait98,  Bogaert94, dolf01, Besk99},  2MASS, MSX,  IRAS, and ISO SWS data,  2 epochs of Broadband Array Spectrograph System spectrophotometry  \citep []{Hackwell90}, {\it Spitzer} IRS data originally published by \citet[]{Juhasz10} ,  AKARI, and WISE data (excepting band 2, 4.5 $\mu$m). We also include the  HST/NICMOS F110W photometry  as well
as the photometry obtained in tandem with the HiCIAO PDI observation and the SIRPOL data.  The data, color-coded by source, are shown in Figure 1, together with a Kurucz T=8250 K photospheric model, reddened by an R=3.1 extinction curve with  E(B-V)=0.13. A similarly good fit at short
wavelengths can be achieved without foreground reddening and with T$_{eff}$=7580K, as used by \citet[]{Andrews11}.  This model requires
 a contribution from UV excess emission (e.g. accretion luminosity) to match the
observed FUV flux  \citep[]{Martin08}, but is consistent with low  line-of-sight extinction  (E(B-V)$\leq$0.1) for
the source to be detected by FUSE.  

\subsection {VLT/NACO Sparse Aperture Masking Interferometry}

MWC 758 was observed on 11 March 2012 with VLT/NACO Sparse-Aperture Masking\footnote[2]{program id: 088.C-0691(A)}.
Observations were taken in the L$'$ band 
($\lambda_{\mathrm{L'}}$= 3.80 $\pm$ 0.31 $\mu$m) using the ``7 holes'' 
aperture mask and the IR wavefront sensor (WFS). A total of 24  data cubes, each one made of 330 frames of 0.12s integration
time were obtained for MWC 758
in L$'$ on the night of 2012-03-12(UT).  Observations of two calibrator
stars , HR 1921 and HD 246369, were obtained for a total of 8 data cubes, interleaved every 8 MWC 758 datacubes.     The MWC 758 observations 
were processed with the Observatoire de Paris SAMP pipelines \citep[]{Lac11a, Lac11b}.
 
The use of the \char'134 7 holes" \citep[C7-892,][]{Tut10}
aperture mask transforms the telescope into a 
Fizeau interferometer. The point spread function is a complex
superposition of fringes at given spatial frequencies. In specific cases,
pupil-masking can outperform more traditional differential imaging
for a number of reasons \citep[][]{Tut06, Lac11a}.
First, the masks are designed to have
nonredundant array configurations that permit phase deconvolution;
slowly moving optical aberrations not corrected by the AO
can be accurately calibrated. Second, the mask primarily rejects
baselines with low spatial frequency and passes proportionately
far more baselines with higher $\lambda$/B 
(where B is baseline length) resolution than does an orthodox
fully filled pupil. Third, high-fidelity recovery of phase
information allows \char'134 super resolution", with a marginal loss of
dynamic range up to $\lambda$/2D (where 
D is the mirror diameter). The principal drawback is a loss
in throughput so that photon and detector noise can affect the
signal-to-noise ratio even where targets are reasonably bright for
the AO system.  The effective field-of-view of SAM
is determined by the shortest baseline so that the technique is
not competitive at separations that are greater than several
times the formal diffraction limit.  For more details on the SAM
mode, please see e.g. \citet[]{Lac11b, Tut10}.

\section {Results}

In this section we discuss the disk detection, compare it with synthetic imagery for a model fitting the sub-millimeter continuum data and the IR SED,  report the 
discovery of spiral arms in the disk, and following \citet[]{Muto12} discuss the implications for a giant planet capable of exciting the more completely imaged
spiral arm. We next discuss limits on stellar, brown dwarf, and giant planetary mass companions provided by the sparse aperture masking data and the
Subaru observations. 

\subsection {MWC 758 as a Gapped  Disk}

Previous studies of MWC 758 had identified this system as hosting a cavity visible in sub-millimeter continuum interferometry, but lacking a very
conspicuous dip in the IR SED near 10$\mu$m \citep [] {Isella10,Andrews11}.  In contrast, our composite SED based on an assembly of archival photometry and spectrophotometry clearly shows  a dip (fig. 1), with indications of mid-IR variability, providing the first independent confirmation that MWC 758 is what has been termed a gapped or \char'134 pre-transitional" disk with an inner disk structure \citep {Isella08}, a wide gap, and an outer disk seen 
in millimeter data \citep[]{Isella10}, and thermal emission \citep[]{Marinas11}.  When compared with other
transitional disks, the depth of the 10 $\mu$m dip is smaller, consistent  with the lack of a cavity in our F110W, H, K$'$ and K$_s$ imagery. 
Modeling of other disks with HiCIAO observations have suggested that the absence of a cavity at H is not atypical of transitional disks, and
can be accounted for by  grain filtration \citep[]{Rice06}, with large mm-sized grains deeply depleted in the cavity region, while smaller grains, which are
more tightly coupled to the gas, are less depleted \citep []{Dong12, Zhu12}.

\subsection {The Structured Disk of MWC 758}

The disk of MWC 758 is detected in polarized, scattered light from 0\farcs2  to 0\farcs 8  (223 AU assuming d=279 pc or 160 AU assuming d=200 pc, fig. 2) 
in polarized intensity at  H-band (fig. 2). The disk detection did not require unusually favorable conditions  \citep[e.g. MWC 480, ]{Kusakabe12}: H-band photometry  was in the range of previous
NIR data for the star  (fig. 1). The detected nebulosity is highly asymmetric about the star, with the greatest extent seen to the west of the star at H, while to the east, nebulosity is not detected exterior to 0\farcs5.  To the west of the star, the scattered light disk extends as far as the sub-millimeter continuum can be traced \citep[]{Isella10} (fig. 3a,b).   The disk is also detected at K$_s$ (fig. 2c) to within 0\farcs1 of the star, and in  K$'$ (fig. 2d) to within 0\farcs25, and at 1.1$\mu$m
(fig. 2e). A false-color composite of the H PI (blue) and K$'$ (red) is shown in figure 2f. This composite image also allows us to identify a region of reduced polarization intensity along PA=155$\pm$5$^\circ$ and PA=330$\pm$5$^\circ$, aligned with the projection of the disk semi-minor axis.  To the east of the star, the H PI  surface brightness $\propto$R$^{-5.7\pm0.1}$ (fig. 3).  To the west of the star,  it drops as  $\propto$R$^{-2.8\pm0.1}$, consistent with the disk exhibiting some flaring. 

Unlike
other protoplanetary \citep[]{Kusakabe12} or transitional disks \citep[]{Muto12} imaged with HiCIAO in H PI, the reflection nebulosity  is not aligned with the disk semi-major axis (PA=65$\pm$7$^\circ$). 
Moreover, we detect scattered light within the region of partial clearing
reported in the sub-millimeter continuum \citep[]{Isella10}, rather than the expected cavity. 
The PI for r$\geq$0\farcs2 is 6.095$\pm$0.085 mJy,  or 0.09\% relative to the star.   Between 0\farcs3 and 0\farcs5 the PI is 2.72 mJy, or 0.0425\% relative 
to the star, indicating that the region of the arms is providing half of the observed polarized intensity. 
If we assume neutral scattering,  the PI in the arm region 0\farcs3$\leq$r$\leq$0\farcs5 can be compared with the F110W total intensity of 0.32\%, 
indicating  a polarized light fraction of $\sim$13.3\% which is in the range of polarization fractions observed for other Herbig Ae stars.

 \subsection {Comparison with Monte-Carlo Radiative Transfer Models for the Disk}
 
 Given the structure seen in the H-band PI imagery,  it is natural to explore what  we would have expected to have seen based on combined
 modeling of the sub-millimeter data and the IR SED. To address this goal, we carried out Monte Carlo radiative transfer modeling  using the code developed by \citet{whi03a,whi03b,whi12}, with modifications as
described in  \citet{Dong12}. In particular, we adopt a disk outer radius of  200 AU, viewed at i=23$^\circ$, and assume an  accretion rate of 
$\dot{M}=10^{-8}M_\odot$ yr$^{-1}$ \citep{Andrews11}.  We model the entire disk with two components: a thick disk with small grains ($\sim\mu$m-sized and smaller), which represents the pristine dust grains from the star forming environment, and a thin disk with large grains (up to $\sim$mm-sized), which is the result of dust growth and settling toward the disk mid plane. We adopt the interstellar medium dust model in \citet{kim94} for our small-grain dust, and the dust model 2 from \citet[]{woo02} for our large-grain dust, which has a power law size distribution up to 1 mm (with power law index 3.5). The large-to-small-dust mass ratio is assumed to be 0.85/0.15 as in \citet[]{Dong12}.  The properties of the grains can be found in \citet{woo02}.

With these assumptions, the total dust mass of the disk is $0.067M_{\rm J}$. The radial surface density distribution of both dust grain populations is assumed to have the following profile
\begin{equation}
\Sigma(R)=\Sigma_{\rm 0}\left(\frac{\rm 1 AU}{R}\right)^\alpha,
\label{eq:sigmasout}
\end{equation}
where $\Sigma_{\rm 0}$ is the normalization factor, $R$ is radius, and $\alpha$ is the power law index. In the vertical direction, we assume
\begin{equation}
\rho(R,z)=\frac{\Sigma(R)}{\sqrt{2\pi}H} e^{-z^2/2H^2},
\label{eq:rhorz}
\end{equation}
where $\rho(R,z)$ is the local volume density, $z$ is the vertical dimension, and $H$ is the scale height (input parameters in the code). Radially, both scale heights for the small and large dust vary with radius as
\begin{equation}
H\propto R^\beta,
\label{eq:b}
\end{equation}
where $\beta$ is a constant power law index, assumed to be 1.08 in this work. The scale height, h, of the small dust is assumed to be 15 AU at 100 AU, while  that for the large dust is assumed to be 1/5 of the $h$ for the small dust, to reflect the fact that big grains tend to settle to the disk mid-plane, while small grains tend to be well coupled with the gas and  have a much larger vertical extension \citep{dul04,dul05,dal06}. 

Our disk model has a gap structure located from 0.5 AU to 73 AU. Inside the gap there are no large  dust grains, as suggested by the spatially resolved SMA observations, which revealed a clear cavity 73 AU in radius at 880 $\mu$m \citep{Andrews11}.  However, the H-band PI data demonstrate that there is a significant amount of scattered light within this radius, indicating the presence of starlight-scattering small dust grains. We set $\alpha=1$ for both dust populations  in the outer disk, and $\alpha=-3$ for the small dust inside the gap, with a continuous surface density distribution across the gap edge. As discussed in detail in \citet{Dong12}, this kind of surface density distribution of the small dust results in a heavy depletion at the inner disk, producing the IR excess depression around 10 $\mu$m on the SED seen in transitional disks, but a smooth NIR scattered light image with no breaks or discontinuities of the surface brightness radial profile at the inner edge of the outer disk.  MWC 758  has a NIR excess  (fig. 1), and VLTI/AMBER data \citep []{Isella08}  indicating a small amount of  hot, small dust
grains  at the inner-most radii. To reproduce these features, we put a inner rim located at 0.33 to 0.5 AU. The surface density of the rim is $\sim0.002$g/cm$^2$. 

Figure 4 shows the model SED and $H$-band polarized light image from the modeling, with the disk major axis horizontal, and the near side of 
the disk in the bottom half of the image.  The model images show both the PI as modeled, and after convolution with an observed HiCIAO PSF(see \citealt{Dong12} for details). The model reproduces the extent of the polarized intensity to the west of the star, but, as expected, predicts a similar
extent of the scattered-light disk to the east of the star, beyond 0\farcs5.  The model predicts a bright arc on the near-side of the image, which is not seen in the data.  The raw model imagery predicts a cavity visible at H in the immediate vicinity of the star, which is not
conspicuous after convolution with the PSF, and which is in the region occulted by the coronagraph in our H-band data.  Higher Strehl ratio and smaller inner working angle
imagery, such as can be provided by  extreme AO systems, will be required to test these predictions of the model.

\subsection {Spiral Arms} 
The most distinctive  features in the NIR images  which are not captured in the  model imagery are  a spiral
feature on the east side of the star, wrapping toward the S, which we term the SE arm,  and a similar feature  originating in the NW and wrapping to the N, which we term the NW arm. The spiral arms  are most clearly detected interior to 0\farcs5  (100 AU for d=200 pc, or 140 AU for d=279 pc). Our  data for MWC 758  therefore demonstrate that SAO 206462 \citep []{Muto12} is not unique in hosting spiral arms straddling the region of the partially cleared millimeter cavity. Moreover,  like SAO 206462, and unlike HD 142527 \citep []{casassus12, rameau12} there is no indication of a cleared zone at H-band for  r$\geq$0\farcs 2 (40 AU for d=200 pc, 56 AU for d=279 AU).  

The spiral arms in the disk of MWC 758 are detected  as a contrast in surface brightness.  The drop in polarized surface brightness immediately exterior to the SE arm is consistent with that arm fully shadowing the outer disk.  Shadowing by the NW arm is less complete: some signal is seen beyond 
the arm in our 2011 data.  We measure the  SB contrast of the arm features by measuring the amplitude at the SE arm  at PA=90$^\circ$ and the corresponding
surface brightness the same distance from the star at PA=270$^\circ$. For the SE arm we find a ratio of  50\%$\pm$20\%. 
The angular extent of the NW arm is sufficiently small that to confidently measure the amplitude and fit the pattern we will require data obtained
with a combination of higher Strehl ratio, and a coronagraph with smaller inner working angle, such as may be provided by SCExAO in tandem
with AO188+HiCIAO on Subaru \citep[]{Martinache11, Martinache12}.  We therefore restrict all but qualitative discussion of the arms to the SE arm.   

Both of the arms have the same rotation sense. If they are trailing structures, as expected for spiral density waves, the rotation of the disk, and 
any companions associated with the arms,  is clockwise. Comparison with the millimeter data \citep[]{Isella10},  then indicates that the northwest side of the disk is the near side of the disk. Our PI image is fully consistent with the disk being viewed at  a  low inclination from face-on,  since it lacks the compression of the PI toward the disk semi-major axis which is conspicuous at i$\geq$38$^\circ$ \citep[]{Kusakabe12}. Given the asymmetric illumination of the disk, we  adopt the millimeter disk inclination i=21$\pm$2$^\circ$  \citep[]{Isella10}. 

\subsection {The disk at other wavelengths}

To be correctly interpreted as spiral density waves,  the spiral arms should be recovered in observations made at different wavelengths and with
different instruments. The IRCS K$'$ data detect the disk in scattered light to at least 0\farcs 5 from the star, and recover both spiral arms using either
classical PSF subtraction and using A-LOCI  as long as we set $dr$ to be larger  than the size of the  disk gap ($dr$ $>$ 25).  
The K$_s$ data sample the disk less completely, due to the very short integration times for each individual exposure. However, they recover
the inner portions of the arms, in the conservative LOCI processing of the data,  to an inner working angle of 0\farcs1 but suppress them in  aggressive
LOCI processing.  

The brightest portion of the disk as seen at 1.1 $\mu$m is the region occupied by the spiral arms, between 0\farcs3 and 0\farcs5. 
 The HST/NICMOS data from 2005 provide no indication of the
asymmetric illumination seen in the 2011 data: this may either be due to changes in illumination over time, or to contrast limits of the HST/NICMOS data. However,
we note that asymmetric illumination of the disk was not seen in the HST/STIS data from 2000. Additional observations will be required to
establish whether the illumination of the outer disk changes with time. The locations of the arms
 coincide with the  brightest 12$\mu$m signal from the MWC 758 disk \citep[]{Marinas11}.  The coincidence of  structural features in
scattered light, polarized scattered light, and in thermal emission indicate that the spiral arms are density features, and not merely distortions
of  the  disk surface. The rapid drop in surface
brightness seen in the H-band PI data to the east of the spiral arms  (fig. 3) further indicates that the arm is  optically thick and extends sufficiently far above
the undistorted disk surface to cast a long shadow.  The region of the NW arm near PA=340$^\circ$
coincides with the clump in continuum emission noted by \citet[]{Isella10}, although with a synthesized beam size of 0\farcs76$\times$0\farcs56, the
available sub-millimeter continuum interferometry lacks  the angular resolution to directly detect the arms. 
As the angular resolution of sub-millimeter interferometry improves, we expect that these features should
be detectable in continuum and gas observations, such as can be provided once ALMA reaches its full extent. 

\subsection {Fitting the  Spiral Arms} 

We follow the approach given in \citet []{Muto12} to fit the spiral arms.  Our initial fits have been restricted to the H-band PI data. 
Given the nearly face-on system inclination, we have not de-projected the data to compensate for the non-zero inclination, but have compensated for the r$^{-2}$ drop in illumination of the
disk (fig. 5). 
 From the data, we have looked for the local maxima of the surface brightness $\times$r$^2$ profile for each radial profile 
 both for the SE spiral and the NW spiral.   If the features are a part of a  ring, the points would lie horizontally in a polar coordinate plot of the
 intensity. The SE spiral is clearly non-axisymmetric, but only a small angular extent of the NW arm is non-axisymmetric, given the inner working
 angle of our data.  For the remainder of this study we restrict
our attention to the SE spiral.  

Assuming i=21$^\circ$ and PA=65$^\circ$ for the disk major axis, the fitting function for the spiral is:
\begin{eqnarray}
 \theta(r) &=& \theta_0 + \dfrac{\mathrm{sgn}(r-r_{\rm c})}{h_{\rm c}}
  \times
  \nonumber \\
 &&  \left[
      \left(\dfrac{r}{r_{\rm c}}\right)^{1+\delta}
      \left\{ \dfrac{1}{1+\delta} - \dfrac{1}{1-\gamma+\delta}
       \left( \dfrac{r}{r_{\rm c}} \right)^{-\gamma} \right\}
      - \left(\dfrac{1}{1+\delta} - \dfrac{1}{1-\gamma+\delta}\right)
     \right]
 \label{spiralform}
\end{eqnarray}
Here, as in SAO 206462 paper, 
The corotation point ($\sim$ launching point of the spiral) is at
$(r_\mathrm{c}, \theta_0)$.  
The disk rotation profile is $\Omega(r) \propto r^{-\gamma}$ 
and the disk sound speed profile is $c(r) \propto r^{-\delta}$.  The disk
aspect ratio ($H/R$) at the corotation radius is $h_c$. Note that assuming $H=c/\Omega$, $\beta$ that appears in Equation 
(3) is related to $\gamma$ and $\delta$ by $\beta=\gamma-\delta$.  We 
also note that if the disk temperature varies as $T \propto r^{-q}$, $q$ 
and $\delta$ are related by $\delta=q/2$ since temperature is 
proportional to the square of the sound speed and the disk sound speed profile 
is $c(r)\propto r^{-\delta}$

Note that the sign after the first $\theta_0$ in the right hand side of
equation (4) %%%%Typo MUTO%%%%1 
is different
from that in  \citet []{Muto12}, due to the counter clockwise
rotation of the SAO 206462 disk.

There are five fitting parameters: 
$(r_{\mathrm c}, \theta_0, h_c, \gamma, \delta)$.
Among them, three $(h_c, \gamma, \delta)$ are determined by the disk
structure and the two $(r_{\mathrm c}, \theta_0)$ determine the location
of the spiral feature. Assuming   Keplerian rotation, $\gamma=1.5$.  In our study of SAO 206462 (Muto et al. 2012), the sound speed profile 
$\delta$ was fixed, but we have varied  this parameter as well. There is a range of values  related to $\delta$ in 
the literature, ranging between 0.45 (the $\psi$ parameter in Andrews et 
al. 2011),  0.19  \cite[q in][]{Chapillon08}, and  0.05  \citep[the $\zeta$ parameter in][]{Isella10}.   We note that for $\delta$=0.5, the opening angle of the disk is almost
constant, in disagreement with our radial SB profile. We looked for the best-fit parameters in the least $\chi^2$ sense by
dividing the $(r_{\rm c}, \theta_0)$ space in $128 \times 128$ cells and
$(h_{\rm c},\delta)$ space in $100 \times 100$ cells with equal spacing
in the linear scale. The  parameter space explored, and best-fit parameters, are shown in table 1. 
Our value for h$_c$ is similar to that derived  by \citet[]{Andrews11}. 
The external perturber fit resulted in  a reduced $\chi^2$ of $0.68$.

%Now we consider the  degeneracy of the parameter space.  The degeneracy of
%parameters of the spiral position $(r_{\rm c},\theta_0)$ and those of
%the disk parameter $(h_{\rm c},\delta)$ are investigated separately.
%First, we consider  the degeneracy of $(r_{\rm c},\theta_0)$.  
%The best fit to the SE spiral arm places the launching point
%of the spiral outside of both where we detect the spiral arm, and 
%at a radius exterior to the dust disk as seen in the sub-millimeter.   However, the
%possibility that the spiral is launched at some inner radii cannot be totally excluded- although the inner radii are outside the contour of
%99\% confidence level, there is a region where $\delta \chi^2$ becomes
%small at inner radii.
%
%Next we consider the $(h_{\rm c}, \delta)$ parameter space.  
%For each fixed values of $(h_{\rm c}, \delta)$, we looked for the set of
%$(r_{\rm c}, \theta_0)$, which gives the minimum $\delta \chi^2$ value.
% Note that $\delta=0$ is a
%constant temperature, flared disk and $\delta=0.5$ is the constant
%opening angle disk.  Although the degeneracy is significant, we favor
%a disk with a rather flat, constant opening angle geometry.  In this case, the spiral can cast a long shadow over the
%outer disk  rather easily, if the spiral has sufficient amplitude, as
%is observed in the H-band PI imagery. 

%%%%%Revision for the parameter degeneracy paragraphs, MUTO%%%%%
{
Now we consider the degeneracy of the parameter space. The degeneracy of
parameters of the spiral position $(r_{\rm c},\theta_0)$ and those of
the disk parameter $(h_c,\delta)$ are investigated separately. The left
panel of Figure 6 shows the degeneracy of $(r_c,\theta_0)$. The best fit
to the SE spiral arm places the launching point of the spiral outside of
both where we detect the spiral arm, and at a radius exterior to the
dust disk as seen in the sub-millimeter.  It might worth pointing out
that there is a range of parameters where the corotation radius $r_c$ is
inside the spiral arms and where the values of $\delta \chi^2$ becomes
small, although these parameters are outside the range of $99\%$
confidence.

In the right panel of Figure 6, we show the degeneracy in the
$(h_c,\delta)$ parameter space.  For each fixed values of
$(h_c,\delta)$, we looked for the set of $(r_c,\theta_0)$, which gives
the minimum $\delta \chi^2$ value. Note that $\delta=0$ is a constant
temperature, flared disk and $\delta=0.5$ is the constant opening angle
disk. Although the degeneracy is significant, we favor a disk with a
rather flat, constant opening angle geometry.  In this case, the spiral
can cast a long shadow over the outer disk rather easily, if the spiral
has sufficient amplitude, as is observed in the H-band PI imagery.  We
also note that the parameters with warmer disk ($h_c \geq 0.1$) are
favored.
}

\subsection {Estimating the Perturber Mass for the SE Spiral} 
If we assume that a perturbing body excites the spiral arm, the mass of the perturber launching the spiral density waves can be estimated independent of knowledge of where it might be located, 
if there are data indicating the relative disk scale height ($h_{\rm c}$),  the mass of the star, and
the amplitude of the spiral pattern as follows:

M$_p$/M$_*$=(Pattern Amplitude) $(h_{\rm c})^{3}$

Adopting 1.8$\pm$0.2 M$_\odot$ for the stellar mass, $h_{\rm c}$=0.18 from the pattern fitting for the SE spiral,   and 
a pattern amplitude of 0.5$\pm$0.2 from the surface 
brightness contrast for the SE arm, we find a perturber mass for the SE pattern of $\sim$5$^{+3 }_{-4}$  M$_J$, consistent with the
perturber being a giant planet, but excluding stellar or brown dwarf mass objects.   This mass estimate is also consistent with 
accretion continuing onto the star.

 \subsection {Contrast limits on Companions to MWC 758 within or near the disk}
 
 The 5-$\sigma$ contrast map from our sparse-aperture masking (SAM) NACO observations is
presented in figure 7a.  At L$'$  band, SAM is sensitive to companions at
separations $\leq$300 mas from the primary. Attained contrasts range from $\Delta$mag(L) =
-4 to $\Delta$mag(L) = -8.  Contrasts were converted to minimum 
detectable companion mass using the models of \citet[]{Bar98, Bar02},
and adopting an age of 3.7 Myr and a distance of 279 pc. Similar limits are achieved for 
5 Myr and a distance of 200 pc. With these assumptions, minimum detectable mass images are shown in figure 7b.  Our SAM
observations are sensitive down to very low mass stellar companions 
($\sim$80 M$_{J}$): we conclude that MWC 758 is a single star.  Thus, our limits for companions
within the region occulted in the HiCIAO H-band PI data are at the low end of the
stellar/ high end of the brown dwarf range. 

Figure 8a shows  S/N maps for the K$_s$  and K$'$ datasets.  Within 3\arcsec ~of the star, the only point source object is the K=16.98 magnitude
background star, which was detected at 37$\sigma$ at K$_s$.   K-band 5-$\sigma$ contrast limits for the K$_s$ data (from 0\farcs1-0\farcs25) and
the K$' $data (r$\geq$0\farcs25) are shown in figure 9.  The deepest imagery in our set of observations is the IRCS K$' $
data which provide 5$\sigma$ contrasts of 1.4$\times$10$^{-4}$ at 0\farcs25, 2.1$\times$10$^{-5}$ at 0\farcs5,
3.3$\times$10$^{-6}$ at 1\arcsec , and 1.12$\times$10$^{-6}$ at 1\farcs55\ (see fig. 8). 

Adopting an
age of 5 Myr and the COND models (Baraffe et al. 2003), these
contrasts correspond to 5 $\sigma$ detection limits of 15-20 M$_{J}$
at 0\farcs25 , 8-10 M$_{J}$ at 0\farcs5, and 3-4  M$_{J}$ at 1\arcsec , 
and near 2 M$_{J}$ at 1\farcs55 for distances of 200 and 279 pc
respectively. These data exclude brown dwarf-mass companions exterior
to 0\farcs5 and are in general agreement with the mass estimates
based on the spiral arm fitting.  Interior to 0\farcs5,  our upper limits are consistent with
at most a low-mass brown dwarf close companions to MWC 758.  
Our mass limits are  consistent with accretion continuing onto the star, as observed, and as predicted 
for disks with  Jovian-mass companions  \citep []{Rice06, Lubow06}.

 \section {Discussion}
 
{\it Confirmation that MWC 758 is a  gapped disk:}
Historically, transitional disks have been defined by the weakness of the IR SED near 10$\mu$m, compared to the excess at longer
wavelengths (and in some systems at shorter wavelengths), in addition to the detection of a wide gap or cleared cavity in sub-millimeter
continuum aperture synthesis imagery. %and possibly also low gas-to-dust ratios (M\'enard et al. 2012XXX). 
 Figure 1 demonstrates the presence of  a dip in the IR SED near 10$\mu$m.  Together with VLTI data indicating an inner dust belt  \citep[]{Isella08},
 and the central cavity and  outer disk seen in millimeter interferometry \citep[]{Isella10}, and thermal emission \citep[]{Marinas11}, the SED data
 indicate that MWC 758 is a transitional disk.  When compared with other
transitional disks, the depth of the 10 $\mu$m dip is smaller, consistent with less complete clearing of the few to  100 AU region. 
However, the  depletion of millimeter-sized grains \citep[]{Andrews11} is similar to SAO 206462.  

{\it Spiral arms are not unique to SAO 206462:} 
 MWC 758 is the third  Herbig Ae star to show spiral arms in the inner parts of the disk,  extending from the region of the sub-millimeter cavity
\citep[]{Isella10} into the outer disk.  Like SAO 206462 and HD 142527 \citep[]{casassus12, rameau12}, the disk is dynamically warm (h$_c \geq$0.1), 
 indicating that spiral arms should be detectable in other warm disks, if Jovian-mass planets are present in the disk, and the region of
 the (sub)millimeter cavity retains at least some small-grain material.  We identify potentially 2 arms in the disk of MWC 758, although we fit only the one with the larger angular coverage. The spiral arms are seen H, K$'$ and K$_s$, and potentially also at 12$\mu$m, demonstrating that they are  visible both in  scattered light and thermal emission.  
 While unresolved, they are also detectable in HST/NICMOS 1.1$\mu$m imagery.  MWC 758 provides the first clear-cut case where an arm is optically thick, and 
 shadows  the outer disk, accounting for the disk non-detection with HST/STIS (IWA=0\farcs5).  The lack of similar shadowing of the west side of the disk suggests that the NW arm 
 has a smaller amplitude, compatible with a lower mass. 
 
 % The overall agreement between the appearance of the
% disk in scattered light  in the late 2011 observations with the lower angular resolution imagery obtained by HST in early 2005, almost a 7
% year baseline, suggests that any Jovian-mass planet in the disk must lie exterior to TBD AU, suggesting that it must lie exterior to the 
% disk imaged by NIR interferometry \citet[]{Isella08}. 

{\it Limitations of sparse-aperture masking in companion searches for Herbig Ae stars: } 
Sparse-aperture masking trades contrast for resolution, realistically
obtaining contrasts of 5-8 mag in the inner 0.1\arcsec.  This is sufficient
to detect brown dwarf and planetary companions to young, solar-like stars
such as LkCa 15  and  T Cha \citep [] {Kraus12, Huelamo11}.  However, at ages $>$3 Myr, contrasts of $>$10 mag are required to
detect exoplanet and low mass ($<$30 M$_{J}$) brown dwarf companions
to young, intermediate-mass stars such as MWC 758. Nonetheless, a close-in
brown dwarf or stellar companion can have a significant impact on disk
properties  \citep[see, e.g.,][]{biller12}. Thus, constraining the
existence of such a companion is key for realistic disk modeling of the 
full system.

 {\it Comparison of mass and luminosity constraints:}
Our mass estimate for the perturber of the SE arm is consistent with the presence 
 of at least one Jovian-mass, but sub-brown dwarf mass planet in this system.  The width of the sub-millimeter cavity and the dip in the IR SED
 are consistent with dynamical clearing by more than one body \citep[]{Zhu11}, as is the presence of an inner disk with  an inclination which is 20$^\circ$ larger than for the outer disk
 \citep[]{Isella08,Isella10}. 
   The spiral arm appears to continue in to 0\farcs1, based on the K$_s$ imagery, but characterizing the spiral 
 requires imagery with higher Strehl ratio and smaller inner working angle than the data  presented in this study. 
 We estimate the mass of the SE arm perturber, based on surface brightness  contrast, for the SE spiral to be 5$^{+3}_{-4}$ M$_J$, consistent with the pattern being launched by a Jovian-mass planet and excluding a stellar or brown dwarf-mass perturber,
 independent of whether the perturber is in the disk or exterior to it.  This is broadly consistent with the SAM constraints excluding a stellar mass
 companion within 0\farcs3 of the star, and ADI contrasts excluding brown dwarf mass bodies in the outer disk.  Our 5$\sigma$ contrast limit
 for an external companion is consistent with our perturber mass estimate, but does not inform on whether the perturber is has a luminosity
 compatible with the COND models, or is under-luminous relative to those models, 
 as predicted for cold-start models of giant planets \citet[]{Spiegel12}.   Higher contrast imagery will be required to constrain the luminosity evolution of any planetary-mass
 bodies in the disk of MWC 758. 
 
 {\it Comparison with SAO 206462:} 
 MWC 758 resembles SAO 206462 in hosting  2 spiral arms which are trailing with respect to the disk rotation, and which are broadly consistent with
 being associated with  Jovian-mass perturbing bodies. If the spiral arms are independent patterns and are associated with giant planets, the presence of 2 arms each
 in SAO 206462 and MWC 758 and at least 4 in HD 142527 supports  predictions that  wide gaps must be cleared by multiple
 planets \citep[]{Zhu11}.  Our fit to the SE  spiral arm suggests a 
 perturber mass $\sim$10$\times$ larger than for the SAO 206462 perturbers, consistent with shadowing of part of the outer disk.  If sufficiently many transitional disks
 with spiral arms can be imaged,  it will ultimately  be possible to compare the giant planet mass function with studies of radial velocity giant planets, and to search for
 differences in the mass function as a function of stellar properties or system age. 
 
 Even at this early stage in the search for giant planets in transitional disks,  the presence of spiral arms 
 in systems with ages ranging from  3.7-5 Myr (MWC 758) to 9$\pm$2 Myr  \citep [SAO206462]{Mueller11} but  otherwise similar levels of large grain depletion 
 \citep[]{Andrews11} places constraints on whether there is a characteristic age for giant planet formation and disk clearing.  In particular, our data
  can be reconciled with  theoretical predictions that  once gaps are opened,  the inner disk should rapidly drain onto the star \citep[]{Ercolano12}, 
 if and only if giant planet formation  is a process with either a
 variable onset and/or duration.   

%%%%%%%%%My Suggestion on the Inner Hole and Planet, MUTO%%%%%%%%%
%%%%%%%%%It is TOTALLY OPTIONAL%%%%%%%%%
%%%%%%%%%If you think this paragraph is irrelevant, just omit%%%%%%%%%
{\it A Planet and Inner Hole Formation:}
{
If the observed spiral structure is attributed to the existence of an
unseen planet, such a planet might not account for the opening of the
inner hole, since the location of the launching point is favored to be
outside the location of the spirals (or the disk).  Another possibility
is that a planet which excites the spiral structures might actually be
scattered from the inner disk, where the giant planet formation timescale is more
rapid.  In this case, the spiral fitting presented in this paper is
relevant if the planet's eccentricity and inclination are well damped.
If it still retains significant eccentricity, the shape of the spiral is
more complicated and also there is a strong time variability within one
orbit of the planet \citep[]{cresswell07, bitsch10}.  In this case, 
the fitting model for the spiral should be
modified. We also note, that a planet external to the disk, if confirmed and found to be on an
orbit indicating scattering from the inner disk, would imply that the disk of MWC 758 {\it must}
host additional giant planets.  }

 {\it  Implications for the future:} 
 MWC 758 has a steadily increasing body of data that indicates that the disk of this Herbig Ae star hosts at least 2  giant planets which are
 dynamically sculpting the disk, both by clearing a wide partial gap in the disk, and by exciting spiral density waves which can be traced in our
 data to 100 -140 AU (depending on d=200 or 279 pc) from the star.  Such waves are expected to provide a means of funneling disk
 material onto and past any planets, until it reaches the star \citep[]{D-RS11}, consistent with both theoretical predictions for the point at which a companion can choke off
 accretion onto the star \citep []{Rice06, Lubow06}  and the presence of on-going  accretion onto the star \citep []{Besk99}. 
 
 Higher Strehl ratio/angular resolution imagery are required to map the NW spiral in this system. Deeper ADI observations are also needed
 to either directly detect planets or to place more stringent luminosity limits on them. Such advances should be possible as the next generation of
 AO systems and coronagraphs commission in the course of the next year.  Further advances are expected in the sub-millimeter: a key assumption made in this study is that the 
 surface brightness  contrast
 of the spiral arms is a reliable proxy for density contrast measurements made at wavelengths where the dust disk is optically thin. Such data 
 with sufficient resolution to separate the arms from the bulk disk are required, and can be obtained once ALMA reaches its full capabilities, in 
 late 2013.

 \acknowledgments {This work, in part, is based on observations made with the NASA/ESA Hubble Space Telescope, obtained at the Space Telescope Science Institute, which is operated by the Association of Universities for Research in Astronomy, Inc., under NASA contract NAS 5-26555. The NICMOS observations are associated with program HST-GO-10177,
 while the STIS imagery is from HST-GTO-8474. The authors thank the  support staff members of the IRTF telescope  for assistance  in obtaining the SED data, and the IR\&D program at The Aerospace Corporation.  This work is partially supported by KAKENHI 22000005 (M.T.), 23103002 (M.H. and M.H.), 23103004 (M.F.), and 24840037 (T.M.), WPI Initiative, MEXT, Japan (E.L.T.), NSF AST 1008440 (C.A.G.) and 1009203 (J.C.), 
and 1009314 (J.P.W.), and NASA NNH06CC28C (M.L.S.) and NNX09AC73G (C.A.G. and M.L.S.). T. C. was supported by a NASA Postdoctoral Fellowship for most of this work.
We wish to thank the anonymous referee for extremely helpful suggestions which have improved the paper.}

\vfill\eject

 \begin{deluxetable}{lccccc}
  \tabletypesize{\footnotesize}
  \tablecaption{Spiral Arm Fitting}
  \tablewidth{0pt} 
  \tablehead{\colhead{Parameter} &\colhead{Search Range}  & \colhead{Best Fit External Perturber}}
\startdata
r$_c$ & $0\farcs05 \leq r_{\rm c} \leq 1\farcs55$ & $r_{\rm c} = 1\farcs55$\\
$\theta$            &  $0 \leq \theta_0 \leq 2\pi$ &$\theta_0 = 1.72 \mathrm{[rad]}$\tablenotemark{a}\\
h$_c$               & $0.05 \leq h_{\rm c} \leq 0.25$ &$h_{\rm c} = 0.182$\\
$\delta$             & $-0.1 \leq \delta \leq 0.6$ &$\delta = 0.6$\\
\enddata 
\tablenotetext{a}{measured from PA=90$^\circ$}
\end{deluxetable}

% \begin {figure}
%  \plotone {fig1_Hashimoto_HPI_vect.eps}
%  \caption { $H$-band polarization vectors are superposed on the $PI$ image before
% subtracting polarized halo (a) and after subtraction (b). 
% The diameter of software masks are 0.$''$4.
% The plotted vectors are binned with 0.09$''$ $\times$ 0.09$''$ and 
% all plotted vectors' lengths are arbitrary for the presentation purpose.
% Histograms in (c) and (d) represent the distributions of angles between 
% polarization vectors and lines from the mask center to the vector position
% at the radial distance from 0.25$"$ to 0.75$"$. The field of view shown is 2\farcs4 on
% a side, and the imagery is oriented N up, E to left.}
% \end{figure} 
 
 %
 % figure 2:   MWC758_SED_rev.eps
 %
 \plotone{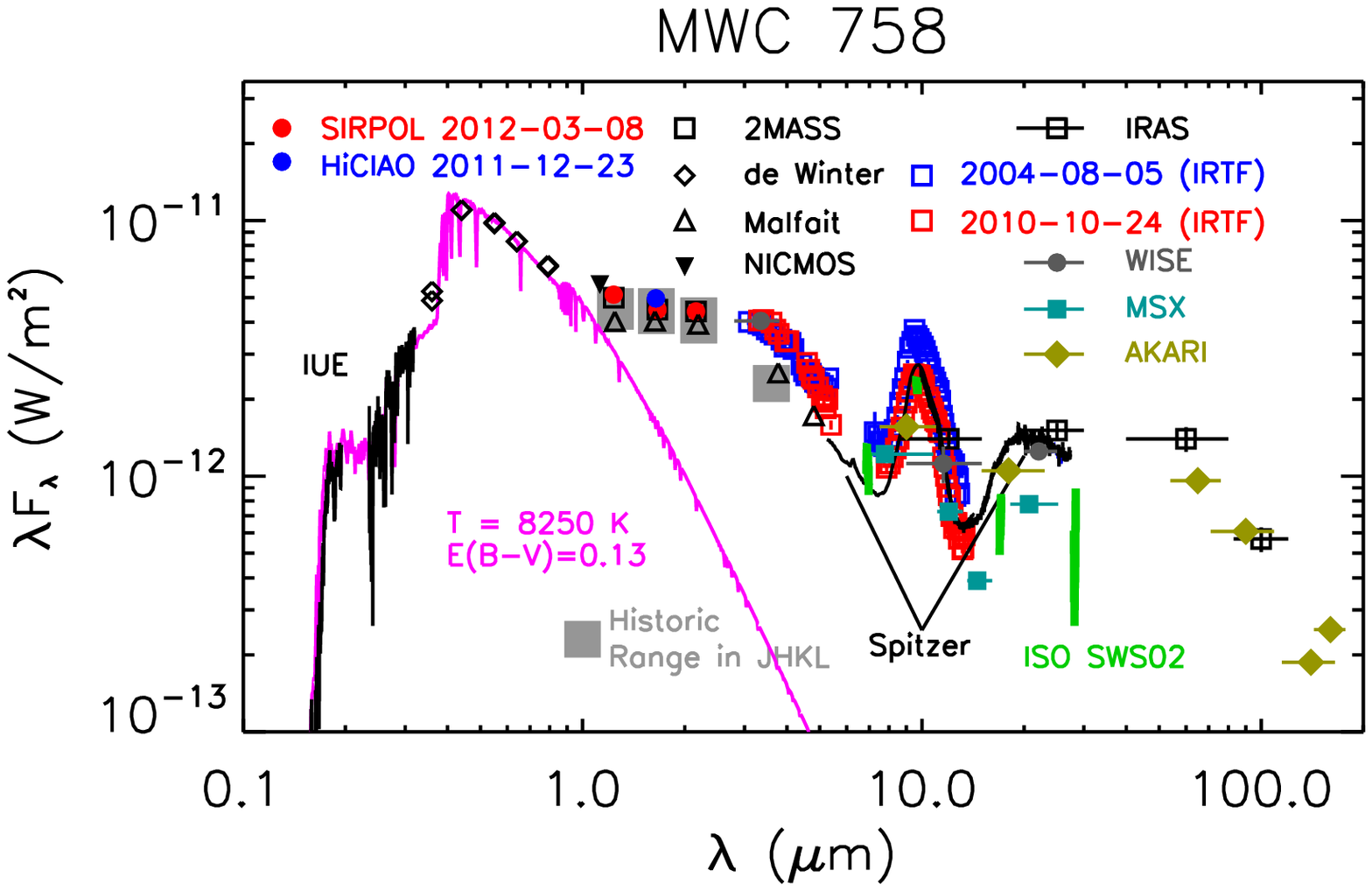}
 \begin {figure}

\caption {Composite SED of MWC 758. The star has limited UV data, optical data indicating minor photometric variability, consistent
with a low inclination, NIR data showing a larger range, and mid-IR data from a variety of sources indicating a factor of 2-3 flux variation
from 5-30$\mu$m.  The mid-IR data, particularly the Spitzer IRS spectrum previously presented by \citet[]{Juhasz10}, clearly demonstrate
the dip in the SED characteristic of pre-transitional, or gapped disks.} 
\end{figure} 
%
%  figure 3:  composite of the disk imagery  mwc758_mosaic_fig2.eps
%
 \begin {figure}
\plotone{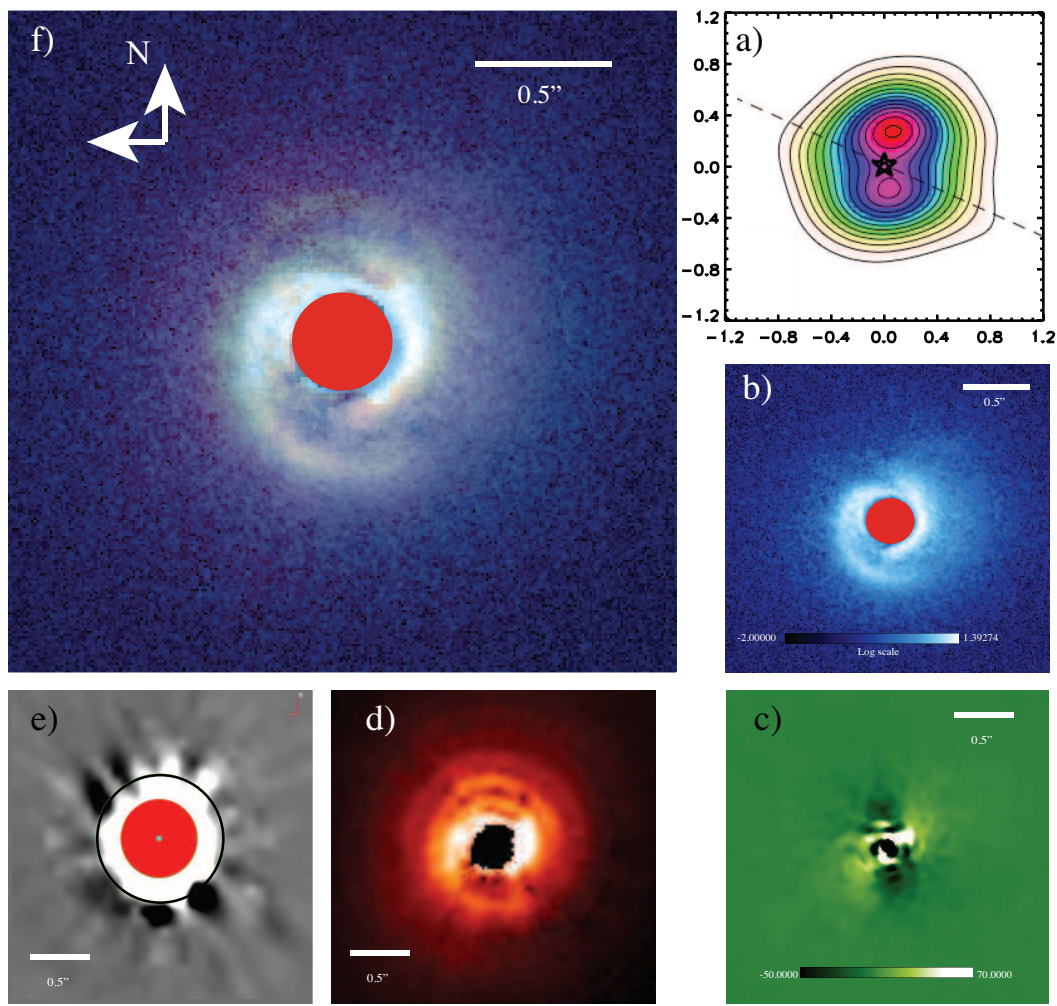}
\caption {The disk of MWC 758: a) the 880$\mu$m continuum after \citet[]{Isella10}, the dashed line indicates the disk semi-major axis,  b) H-band polarized intensity, 
c) K$_s$ data with conservative LOCI processing, d) K$'$ intensity with conservative LOCI processing, e) HST/NICMOS
total intensity data following PSF subtraction, f) color composite of H PI and K$'$ data.    The images are all 2\farcs4 on a side, and oriented with north up and east to the left. The HiCIAO H PI data use an 0\farcs3 diameter coronagraphic spot, the NICMOS data uses an 0\farcs6 spot. The IRCS
data use an 0\farcs15 diameter occulting spot, and the K$_s$ data have an inner working angle of 0.1\arcsec .}
 \end {figure}
%
% figure 3: power law fits east and west  mwc758eastwestradprof_multifit_v8.ps
%
 \begin {figure}
 \plotone{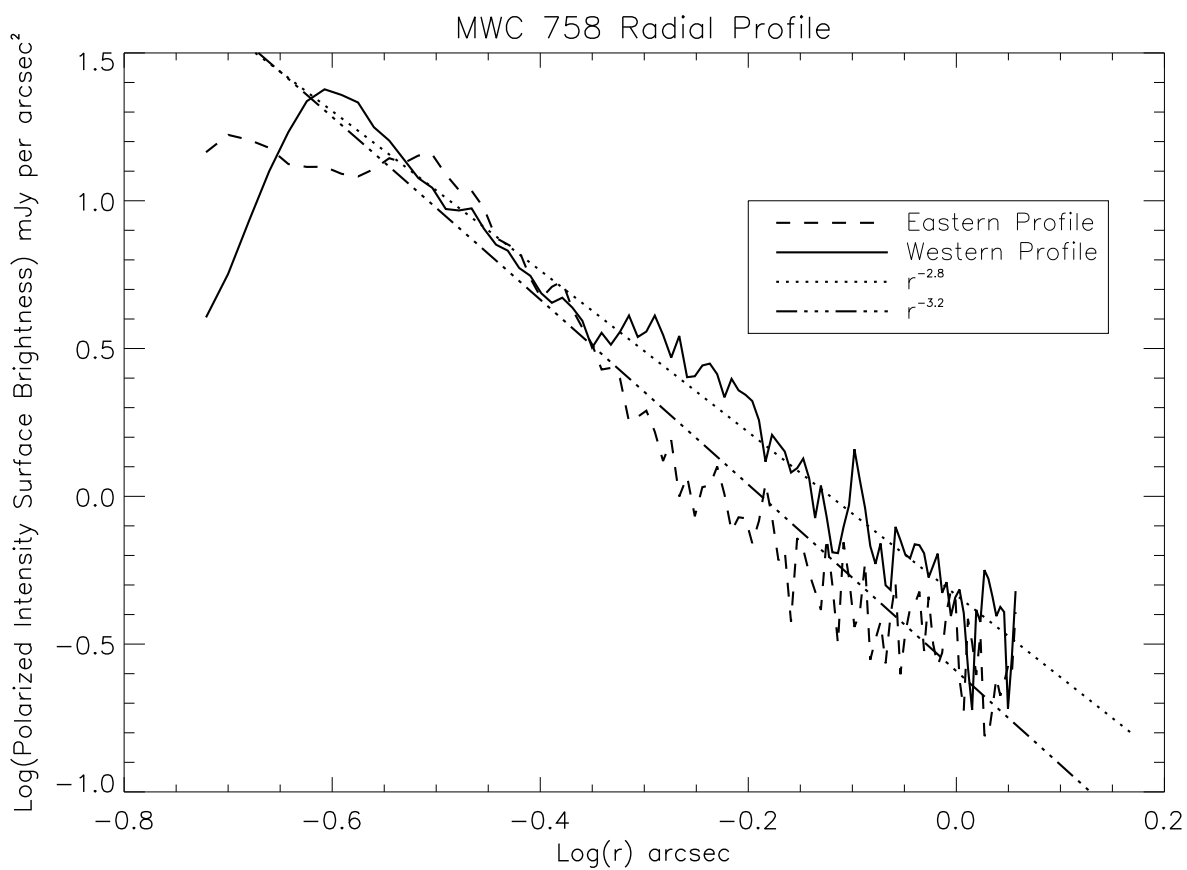}
 \caption{ Radial surface brightness profiles at PA=270$^\circ$ (black) and PA=90$^\circ$(gray) together with power-law fits.}
 \end{figure}

%
%  figure 4: MCRT fit to SED and HPI image  MWC758_model_sed.eps MWC758_model_image.eps
%
\begin{figure}[tb]
%\vspace*{-0.5cm}

\plottwo {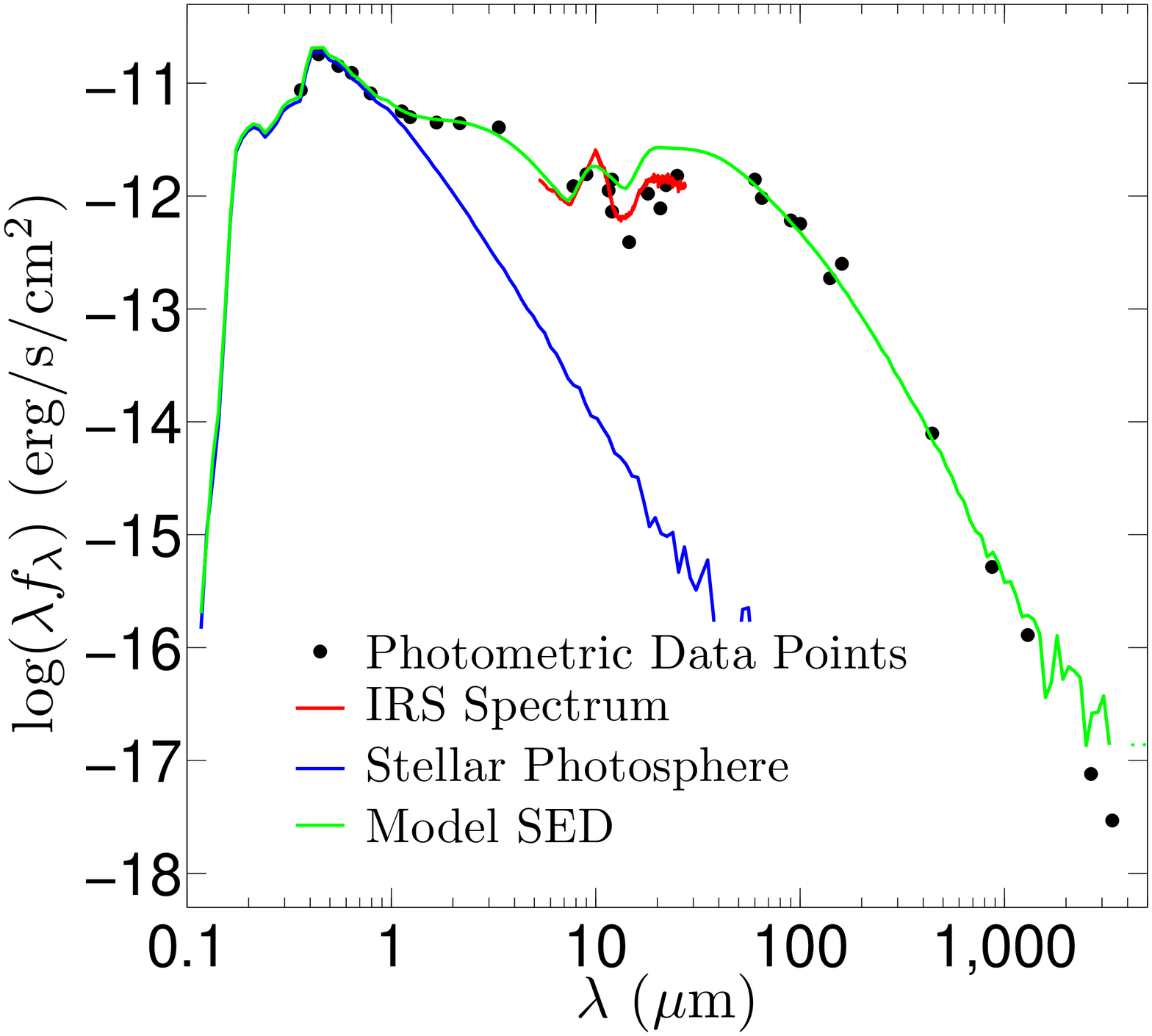}{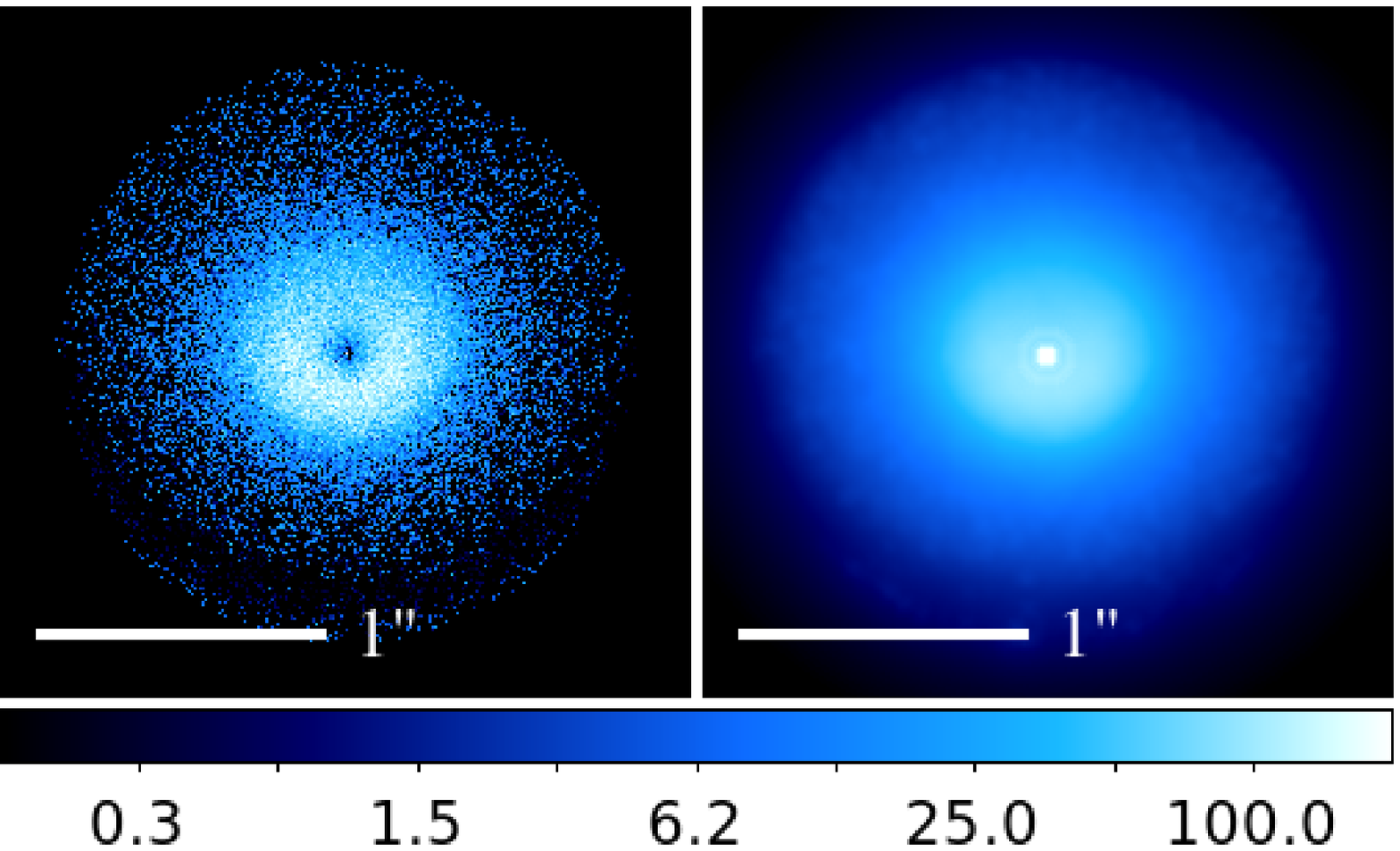} 
%\vspace*{-0.5cm}

\caption{Radiative transfer modeling results for MWC 758, showing the SED (left), the raw model $H$-band polarized light image (middle)
with a cavity visible in the inner disk, and the convolved $H$-band polarized light image (right). The images are in unit of mJy/arcsec$^2$.
\label{fig:modeling}}
\end{figure}
%
%  figure 5: spiral arm fitting spiral_MWC758_small_v2.eps
%
\begin{figure}
\plotone{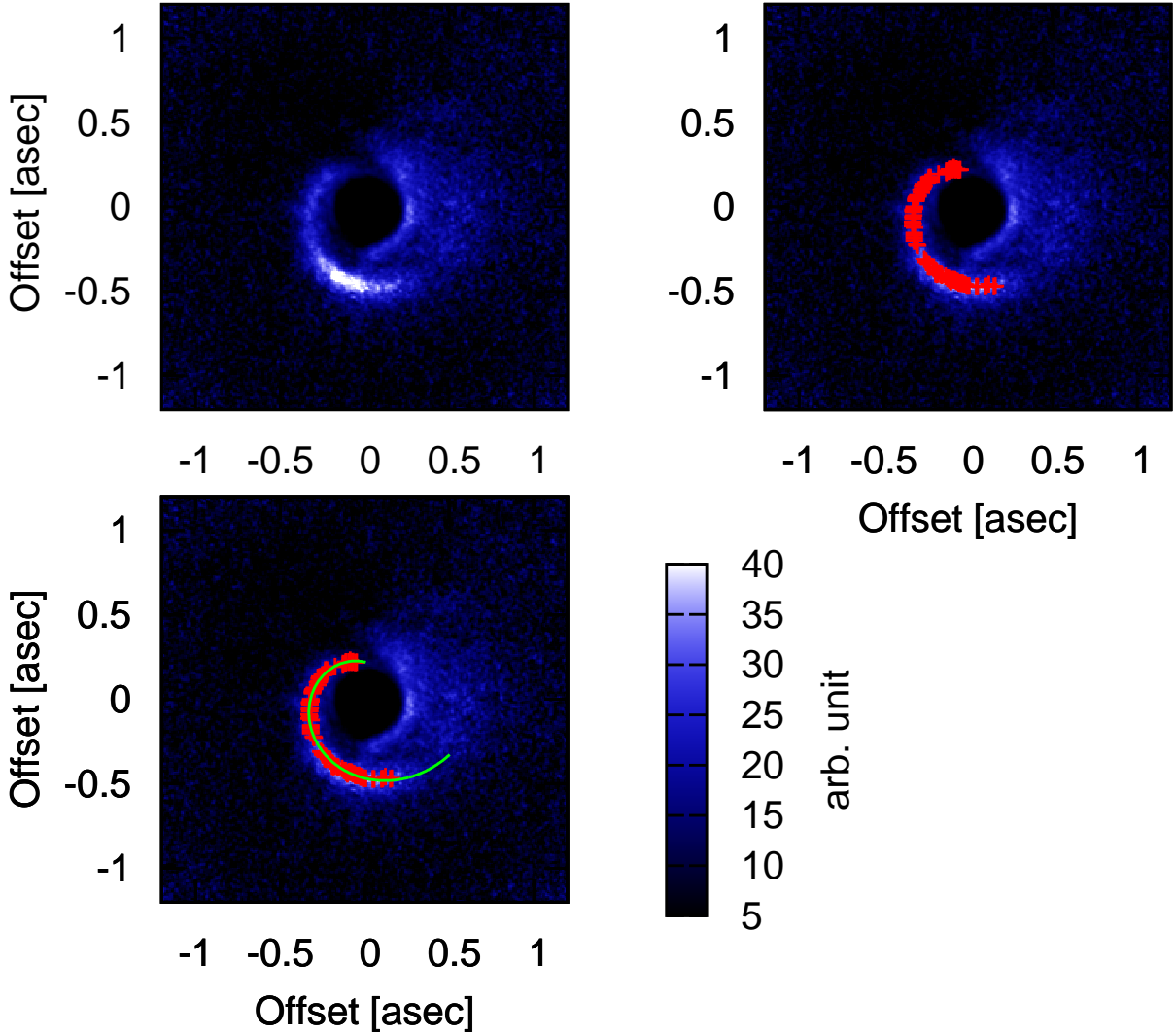}

%\caption {Fitting the SE spiral arm:  top left) PI scaled by r$^2$ to compensate for the radial drop off in illumination
%by the star, top right) Red X's mark the peak intensity along the arm that we have used in fitting, lower left) the
%fit to the peak intensity along the arm is shown in green,  lower right) reduced chi-sqr for the fit. Our best fit
%is for an external perturber (see table 1).} 

%%%%%%%%Revision of Caption for Figure 5, MUTO%%%%%%%%%
%%%%%%%%Figure 5 file name: spiral_MWC758_v3.eps
%%%%%%%%or use spiral_MWC758_v3_small.eps for small file size
%%%%%%%%Download from: http://www.ns.kogakuin.ac.jp/~ft13389/MWC758/revise/ 
\caption{ { Top left: the surface brightness of the scattered light
 normalized by $r^2$, where $r$ is the distance from the central star.
 The shape of the spiral is more enhanced, and this data is used for the
 analysis for the spiral fitting.  Top right: the points that represent
 the spiral shape is shown by red points, which are used as data points
 for the fitting.  Bottom left: the shape of the spiral given by
 Equation (4) with the best fit parameters (Table 1) is shown by green
 curves.  The red points are the same with those in the top right
 panel.} }

\end {figure} 

%
%  NEW Figure 6: Parameter Degeneracy, MUTO
%Figure 6 file name: fit_degen_MWC758_v3.eps
%Download from: http://www.ns.kogakuin.ac.jp/~ft13389/MWC758/revise/ 
\begin{figure}
\plotone{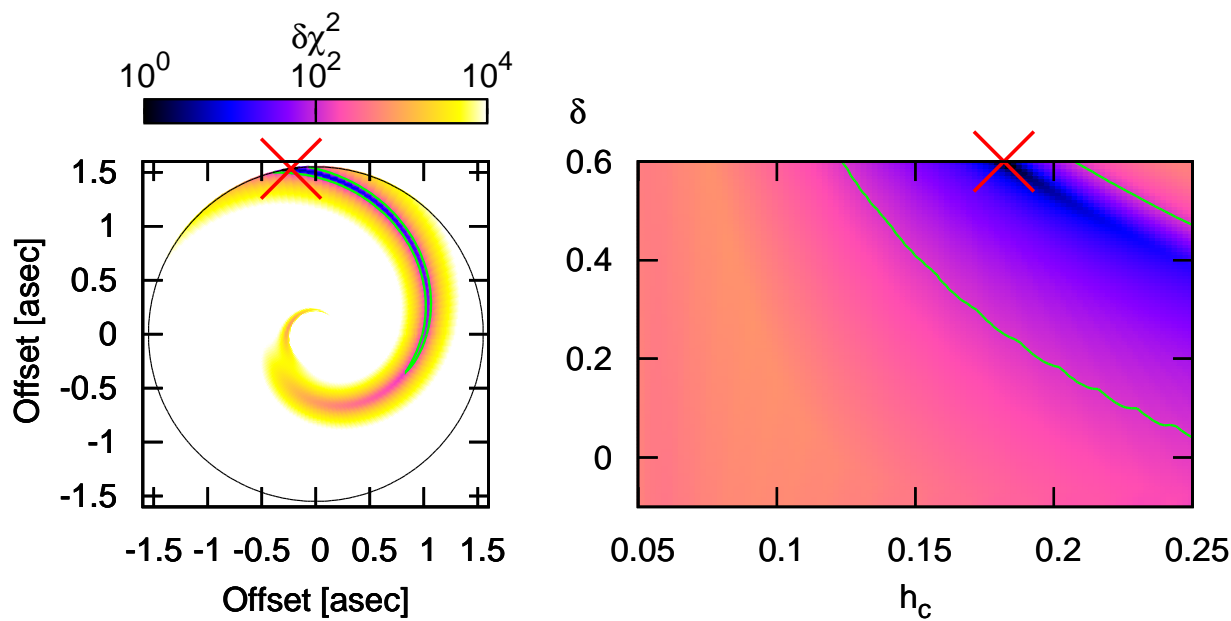}
\caption{ { Left: The degeneracy of $(r_c,\theta_0)$ parameters
 plotted in the 2D spherical coordinate.  The color shows the values of
 $\delta \chi^2$ with respect to the best-fit values, which are shown by
 the red cross.  The contour of 99\% confidence level is shown by the
 green line, and the black circle is the radius of $r=1\farcs 55$, which
 is the upper limit of the parameters that are searched.  Right: The
 degeneracy of $(h_c,\delta)$ parameters.  As in the left panel, the
 contour of 99\% confidence level is shown by green curves and the
 best-fit parameter is indicated by the red cross.} }
\end {figure}

 %
 %  figure 6: SAM artwork MWC758_fig4.eps
 %
 \begin {figure}
 \plotone {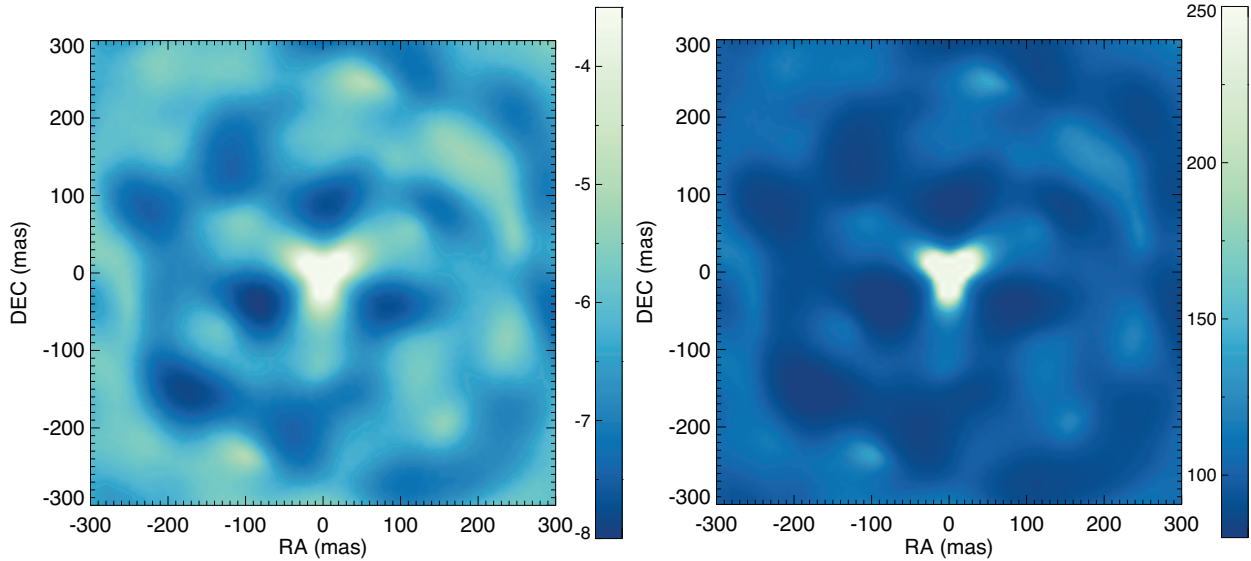}
 \caption {left: 5-$\sigma$ contrast map from our SAM NACO observations.  Our
observations are sensitive to companions 4 to 8 magnitudes fainter
than the primary star at separation of 300 mas or less.
right: Minimum detectable companion mass map.  Contrasts were converted to minimum 
detectable companion mass using the models of \citet[]{Bar98, Bar02},
and adopting an age of 3.7 Myr and a distance of 279 pc.
We are sensitive to companions right down to the star / brown dwarf 
boundary, i.e. $\sim$80 M$_{Jup}$.  From these observations, we
conclude that MWC 758 does not have a low
mass stellar companion within 300 mas of the primary star.
 }
 \end {figure} 
 %
 %  figure 7: SNR maps for aggressive LOCI  MWC758_SN.eps MWC758_ircs_snmap_2
 %
 \begin {figure} 
 \plottwo {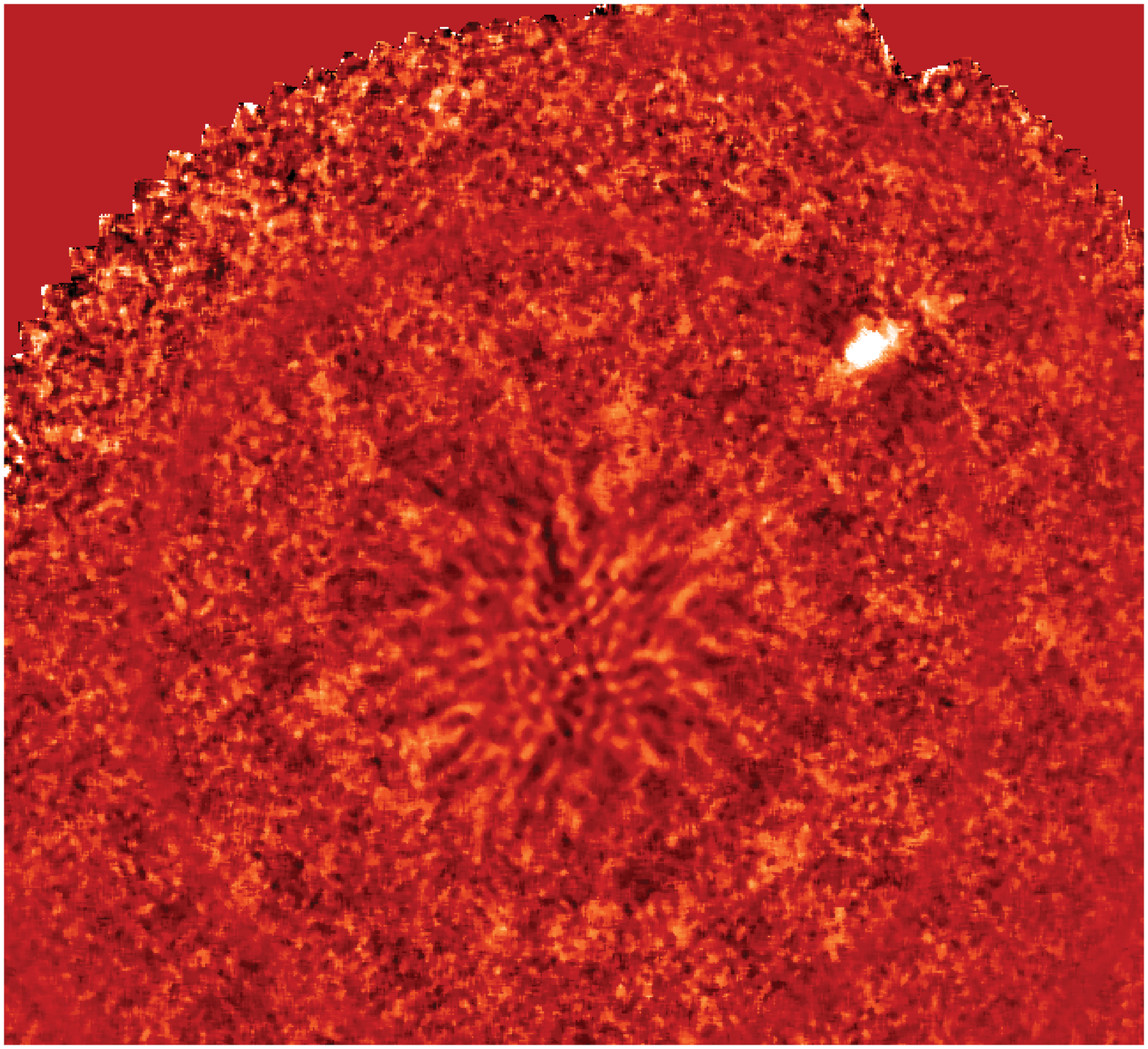}{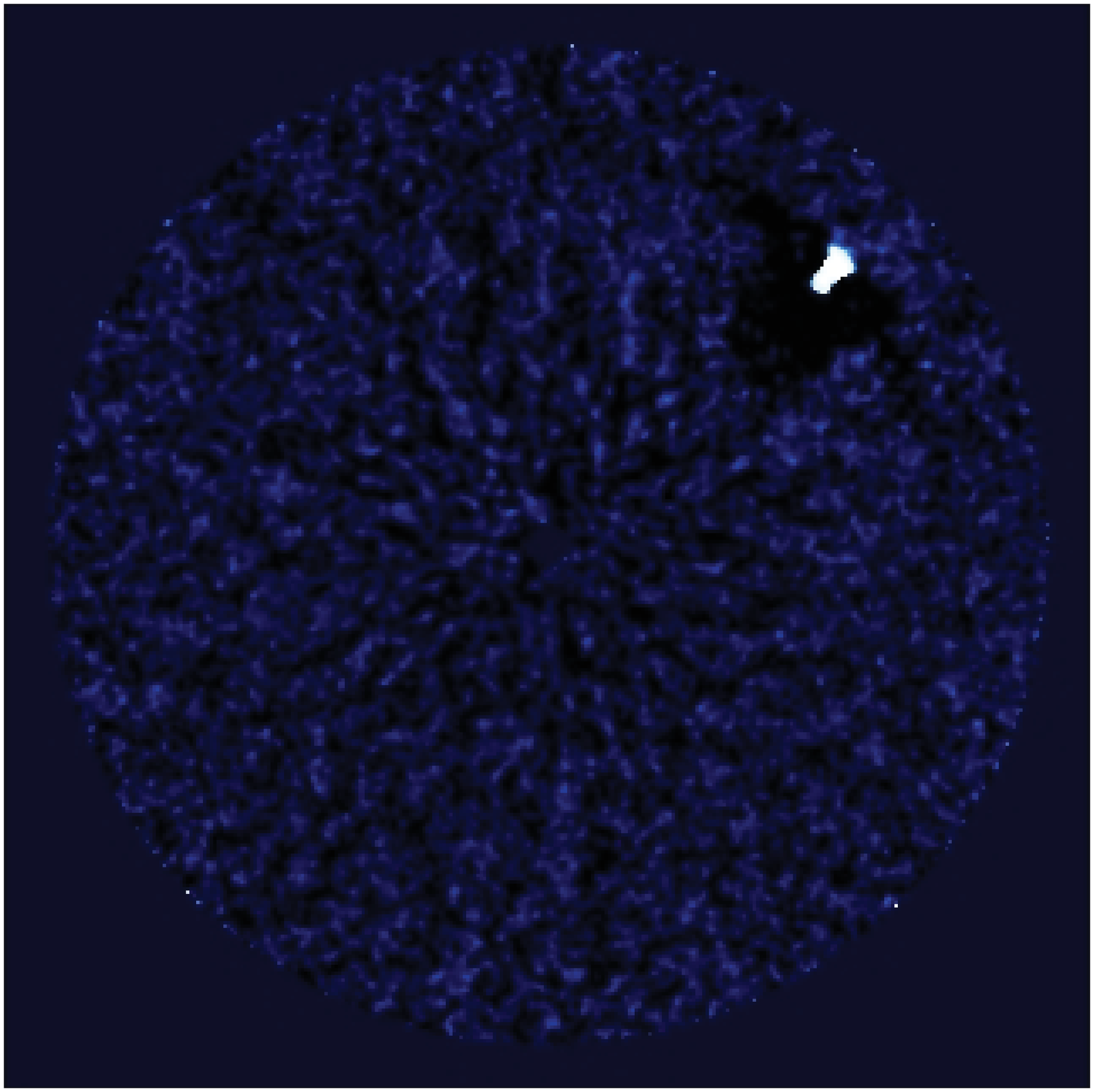}
 \caption {Aggressive LOCI S/N map for MWC 758 at K$_s$ (left) and K$' $(right).  The field shown for the K$_s$ data is $\sim$2\farcs5, while that for 
 the K$'$ data has an outer radius of 3\farcs04. 
  The only point source detected in either dataset is the background object first seen by HST. } 
 \end {figure} 
 %
 % figure 8; 5 sigma contrasts MWC758_Kband_contrast.eps
 %
 \begin {figure} 
\plotone {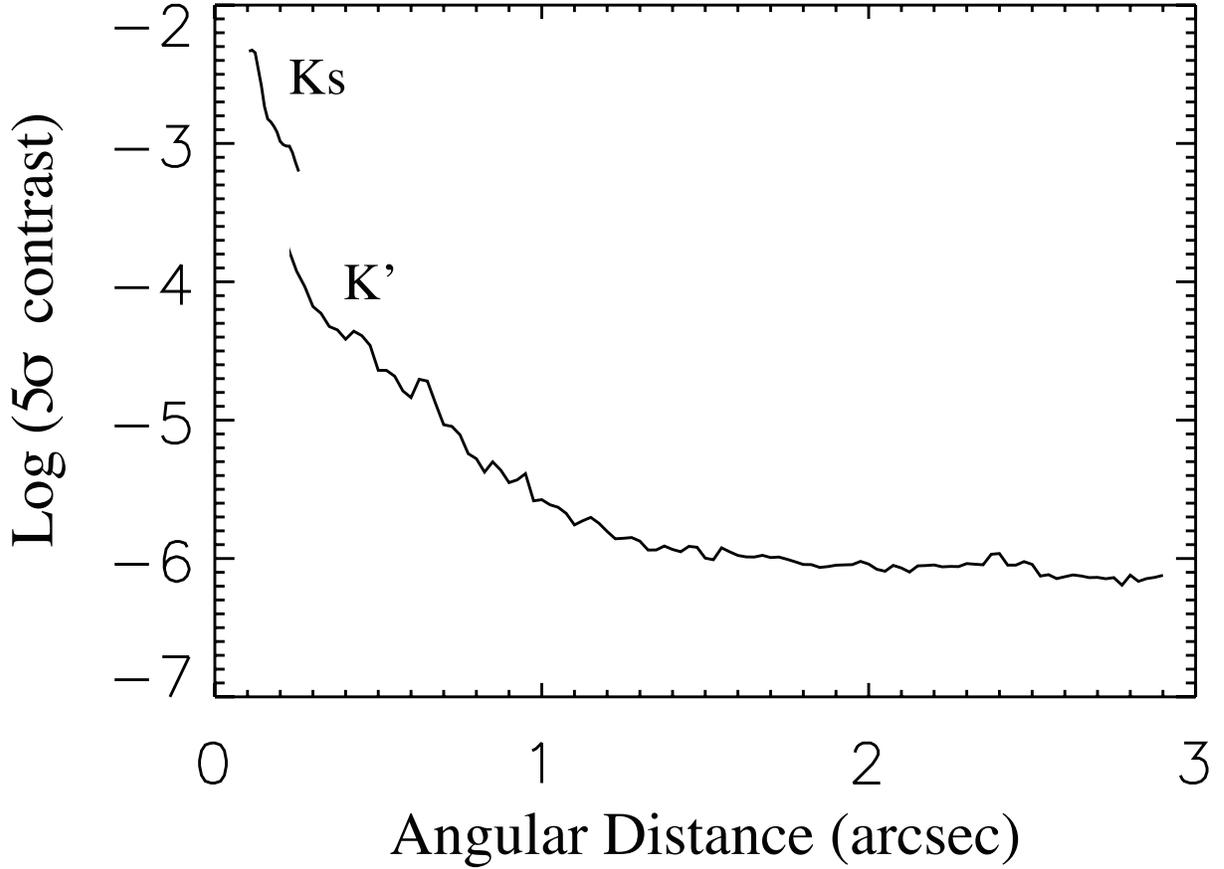}
 \caption { Five $\sigma$ contrast limits for Jovian-mass bodies in or near the disk of MWC 758 derived from the K$_s$ data (0.1-0.25\arcsec ),
 and  the K$'$ data (r$\geq$0.25\arcsec ). The discontinuity at 0\farcs25 reflects the change in exposure depth between the K$_s$ and K$'$ 
 datasets. }
 \end {figure}


\begin{thebibliography}{}
\bibitem[Andrews et al.(2011)]{Andrews11}Andrews, S., et al., 2011, \apj, 732, 42
\bibitem[Baraffe et al.(1998)]{Bar98} Baraffe, I., Chabrier, G., Allard, F., \& Hauschildt, P.~H.\ 1998, \aap , 337, 403
\bibitem[Baraffe et al.(2002)]{Bar02} Baraffe, I., Chabrier, G., Allard, F., \& Hauschildt, P.~H.\ 2002, \aap , 382, 563
\bibitem[Barraffe et al.(2003)]{Barraffe03}Barraffe, I., Charbrier, G., Barman, T.S., Allard, F., Hauschilt, P.H., 2003, \aap , 402, 701
\bibitem[Beskrovnaya et al.(1999)]{Besk99}Beskrovnaya, N.G.. Pogodin, M.A., Miroshnichenko, A.S.,et al. 1999, \aap ,  343, 163
\bibitem[Biller et al.(2012)]{biller12} Biller, B., et al. 2012 \apj , 753, L38
\bibitem[Bitsch \& Kley(2010)]{bitsch10} Bitsch, B., \& Kley, W., A\&A, 523, id.A30 (2010)
\bibitem[Bogaert et al.(1994)]{Bogaert94}Bogaert, E. 1994, \char'134 Multispectrale studie van de aard en de veranderlijkheid van optisch-heldere sterren met circumstellaire stofschillen", Ph.D. Thesis, Kathol. Univ. Leuven, 1994 PHDT....226B
\bibitem[Casassus et al.(2012)]{casassus12}Casassus,  S.,  Perez,  M., S., Jord\'an, A., M\'enard, F.,  Cuadra, J., Schreiber, M. R., Hales, A. S., Ercolano, B. 2012, \apj , 754, L31C
\bibitem[Castillo-Rogez et al.(2009)]{C-R09}Castillo-Rogez, J., Johnson, T.V., Lee, M.H., Turner, N.J. Matson, D.L. \& Lunine, J. 2009, Icarus, 204, 658
\bibitem[Chapillon et al.(2008)]{Chapillon08}Chapillon, E., Guilloteau, S., Dutrey, A., Pi\'etu, V., 2008, \aap , 488, 565
\bibitem[Cresswell et al.(2007)]{cresswell07}Cresswell, P., Dirksen, G., Kley, W., \& Nelson, R. P. 2007, A\&A, 473, 329
\bibitem[Currie et al.(2011a)]{Currie2011a}Currie, T., et al., 2011a, \apj , 729, 128
\bibitem[Currie et al.(2011b)]{Currie2011b}Currie, T., et al., 2011b, \apj , 736, L33
\bibitem[Currie et al.(2012)]{Currie2012}Currie, T., et al., 2012, \apj, 760, L32
\bibitem[D'Alessio et al.(2006)]{dal06} D'Alessio, P., Calvet, N., Hartmann, L., Franco-Hern{\'a}ndez, R., \& Serv{\'{\i}}n, H.\ 2006, \apj , 638, 314
\bibitem[de Winter et al.(2001)]{dolf01}de Winter, D., van den Ancker, M.E., Maria, A., Th\'{e}, P.S., Tjin A Djie, H.R.E., Redondo, I., Eirola, C., \& Molster, F. 2001 \aap , 380, 609
\bibitem[Dodson-Robinson \& Salyk(2011)]{D-RS11}Dodson-Robinson, S.E., \& Salyk, C. 2011, \apj , 738, 131
\bibitem[Dong et al.(2012)]{Dong12}Dong, R. et al. 2012,  \apj , 750, 161
\bibitem[Dullemond  \& Dominik(2004b)]{dul04} Dullemond, C.~P., \& Dominik, C.\ 2004, \aap, 421, 1075
\bibitem[Dullemond \& Dominik(2005)]{dul05} Dullemond, C.~P., \& Dominik, C.\ 2005, \aap, 434, 971
\bibitem[Ercolano, \& Koepferl(2012)]{Ercolano12} Ercolano, B., \& Koepferl, C.  2012, arXiv:1208.4689
\bibitem[Grady et al.(2005)]{Grady05}Grady, C.A. et al. 2005, \apj , 630, 958 %13 authors
\bibitem[Hackwell et al.(1990)]{Hackwell90} Hackwell, J.H., Warre, D.W., Chatelain, M.A., Dotan, Y., Li, P.H., Lynch, D.K., Mabry, D.J., Russell, R.W., \& Young, R.M. 1990, Proc. SPIE, 1235, 171.\bibitem[Hashimoto et al.(2011)]{hash11} Hashimoto, J. et al. 2011, \apj, 729, L17 
\bibitem[Hodapp et al.(2008)]{Hodapp08}Hodapp, K.W., et al. 2008, SPIE 7014, E42H
\bibitem[Hu\'elamo, et al.(2011)]{Huelamo11} Hu\'elamo, N., et al. 2011, \aap , 528, L7
\bibitem[Isella et al.(2008)]{Isella08} Isella, A., Tatulli, E., Natta, A., Testi, L., 2008, \aap , 483, L13
\bibitem[Isella  et al.(2010)]{Isella10} Isella, A., Natta, A., Wilner, D., Carpenter, J.M., Testi, L.  2010, \apj , 725, 1735
\bibitem[Juh\'{a}sz et al.(2010)]{Juhasz10} Juh\'{a}sz, A, Bouwman, J., Henning, Th., Acke, B., van den Ancker, M.E., Meeus, G., Dominik, C., Min, M., Tielens, A.G.G.M., \& Waters, L.B.F.M. 2010, \apj, 721, 43. 1
\bibitem[Kandori et al.(2006)]{Kandori06} Kandori, R., et al. 2006 SPIE 6269E.159K 
\bibitem[Kim et al.(1994)]{kim94} Kim, S.-H., Martin, P.~G., \& Hendry, P.~D.\ 1994, \apj, 422, 164
\bibitem[Kraus \& Ireland(2012)]{Kraus12}Kraus, A. \& Ireland, M.J. 2012, \apj, 745, 5 
\bibitem[Kusakabe et al.(2012)]{Kusakabe12}Kusakabe, Nb. et al. 2012, \apj,  753, 153
\bibitem[Lacour et al.(2011a)]{Lac11a} Lacour, S., Tuthill, P., Ireland, M., Amico, P., \& Girard, J.\ 2011a, The Messenger, 146, 18
\bibitem[Lacour et al.(2011b)]{Lac11b} Lacour, S., Tuthill, P., Amico, P., et al.\ 2011b, \aap, 532, A72
\bibitem[Lafreni\`ere et al.(2007)]{Lafreniere2007}Lafreni\'ere, D., et al., 2007, \apj, 660, 770
\bibitem[Lowrance et al.(2005)]{Lowrance05}Lowrance, P.J. et al. 2005, AJ,  130, 1845
\bibitem[Lubow \& D'Angelo(2006)]{Lubow06}Lubow, S.H. \& D'Angelo, G., 2006, \apj, 641,526
\bibitem[Malfait et al.(1998)]{Malfait98} Malfait, K., Bogaert, E., \& Waelkens, C. 1998, \aap, 331, 211
\bibitem[Mari\~nas et al.(2011)]{Marinas11}Mari\~nas, N. et al. 2011, \apj,  737, 57
\bibitem[Marois et al.(2006)]{Marois2006}Marois, C., et al., 2006, \apj, 641, 556
\bibitem[Martinache et al.(2011)]{Martinache11}Martinache, F., Guyon, O., Garrel, V., Clergeon, C., Groff, T., Stewart, P., Russel, R., Blain, C. 
2011, SPIE. 8151E..22M 
\bibitem[Martinache et al.(2012)]{Martinache12}Martinache, F., Guyon, O., Clergeon, C., Blain, C. 2012, arXiv:1206.2996 
\bibitem[Martin-Za\"idi et al.(2008)]{Martin08}Martin-Za\"idi, C., Deleuil, M., Le Bourlot, J., Bouret, J.-C., Roberge, A., Dullemond, C.P., Testi, L.,
Feldman, P.D., Lecavelier Des Etangs, A., Vidal-Madjar, A., 2008, \aap,  484, 225
\bibitem[Meeus et al.(2012)]{Meeus12}Meeus, G. et al. 2012, \aap, 544, A78M 
\bibitem[Minowa et al.(2010)]{Minowa10}Minowa, Y. et al. 2010, SPIE, 7736, E122
\bibitem[M\"uller et al.(2011)]{Mueller11}M\"uller, A., van den Ancker, M.E., Launhardt, R., Pott, J.-U., Fedele, D., Henning, Th. 2011, A\&A 530,
A85M
\bibitem[Muto et al.(2012)]{Muto12}Muto, T. et al. 2012, \apj, 748, L22
\bibitem[Pereyra et al.(2009)]{Pereyra09}Pereyra, A., Girart, J.M., Magalh\~aes, A.M., Rodriques, C.V., and F.X. de Ara\'ujo, 2009, A\&A 501, 595
\bibitem[Rameau et al.(2012)]{rameau12}Rameau, J., Chauvin, G., Lagrange, A.-M., Thebault, P., Milli, J., Girard, J.H., Bonnefoy, M. A\&A 546A, 24R
\bibitem[Rice et al.(2006)]{Rice06}Rice, W.K.M., Armitage, P.J., Wood, K., Lodato, G. 2006, MNRAS, 373, 1619
\bibitem[Salyk et al.(2011)]{Salyk11}Salyk, C., Blake, G.A., Boogert, A.C.A., Brown, J.M.  2011, \apj , 743, 112
\bibitem[Schneider et al.(2006)]{Schneider06}Schneider, G. et al., 2006, \apj, 650, 414
\bibitem[Spiegel \& Burrows(2012)]{Spiegel12}Spiegel,D., \& Burrows, A. 2012, \apj,  745, 174S
\bibitem[Suzuki, et al.(2010)]{Suzuki10} Suzuki, R. et al. 2010, SPIE 7735, E101
\bibitem[Tamura et al.(2006)]{Tamura06}Tamura, M. et al. 2006, SPIE 6269, E28T
\bibitem[Tamura(2009)]{Tamura09} Tamura, M. 2009, AIPC 1158, 11
\bibitem[Thalmann et al.(2010)]{Thalmann2010}Thalmann, C., et al., 2010, \apj, 718, L87
\bibitem[Tokunaga et al.(1998)]{Tokunaga1998}Tokunaga, A., et al., 1998, SPIE, 3354, 512
\bibitem[Tuthill et al.(2006)]{Tut06} Tuthill, P., Lloyd, J., Ireland, M., et al.\ 2006, \procspie, 6272,  103T
\bibitem[Tuthill et al.(2010)]{Tut10} Tuthill, P., Lacour, S., Amico, P., et al.\ 2010, \procspie, 7735,  56T
\bibitem[van den Ancker et al.(1998)]{vandena98}van den Ancker, M.E., de Winter, D., Tjin a Djie, H.R.E. 1998, \aap,  330, 145V 
\bibitem[van Leeuwen(2007)]{vanLeeuwen07} van Leeuwen, F. 2007 \aap 474, 653
\bibitem[Whitney et al.(2003a)]{whi03a} Whitney, B.~A., Wood, K., Bjorkman, J.~E., \& Wolff, M.~J.\ 2003, \apj, 591, 1049 
\bibitem[Whitney et al.(2003b)]{whi03b} Whitney, B.~A., Wood, K., Bjorkman, J.~E., \& Cohen, M.\ 2003, \apj, 598, 1079 
\bibitem[Whitney et al.(2012)]{whi12} Whitney, B.~A. et al. 2012, in prep.
\bibitem[Wood et al.(2002)]{woo02} Wood, K., Wolff, M.~J., Bjorkman, J.~E., \& Whitney, B.\ 2002, \apj, 564, 887 
\bibitem[Zhu et al.(2011)]{Zhu11}Zhu, Z., Nelson, R.P., Hartmann, L., Espaillat, C., Calvet, N. 2011, \apj,  729, 47Z
\bibitem[Zhu et al.(2012)]{Zhu12}Zhu, Z., Nelson, R.P., Dong, R., Espaillat, C., Hartmann, L., 2012, \apj, 755, 6Z
\end{thebibliography}
\end{document}